\documentclass[manuscript]{acmart}
\AtBeginDocument{%
  \providecommand\BibTeX{{%
    \normalfont B\kern-0.5em{\scshape i\kern-0.25em b}\kern-0.8em\TeX}}}
    
\usepackage{subcaption}
\usepackage{float}
\usepackage{natbib}
\usepackage{graphicx}
\usepackage{rotating}
\usepackage{lscape}
\usepackage{xspace}

\let\oldciteauthor=\citeauthor
\def\citeauthor#1{{\hypersetup{citecolor=black}\oldciteauthor{#1}}}

\copyrightyear{2024}
\acmYear{2024}
\setcopyright{rightsretained}
\acmConference[FAccT '24]{The 2024 ACM Conference on Fairness, Accountability, and Transparency}{June 3--6, 2024}{Rio de Janeiro, Brazil}
\acmBooktitle{The 2024 ACM Conference on Fairness, Accountability, and Transparency (FAccT '24), June 3--6, 2024, Rio de Janeiro, Brazil}\acmDOI{10.1145/3630106.3658984}
\acmISBN{979-8-4007-0450-5/24/06}

\usepackage{pgfplots}
\pgfplotsset{width=10cm,compat=1.9}

\usepackage{stackengine}
\usepackage{verbatim}
\usepackage{xspace}

\usepackage{xcolor}
\definecolor{darkgreen}{rgb}{0.05,0.5,0.05}

\newcommand{\gh}{GitHub\xspace}

\begin{document}

%%
%% The "title" command has an optional parameter,
%% allowing the author to define a "short title" to be used in page headers.
\title[Trust in AI-powered Code Generation Tools]{Investigating and Designing for Trust in AI-powered Code Generation Tools}
% \title{Understanding and designing for developers' trust-building process with AI-powered code generation tools}

%Investigating and Designing for Trust in AI-powered Code Generation Tools

%
% The "author" command and its associated commands are used to define
% the authors and their affiliations.
% Of note is the shared affiliation of the first two authors, and the
% "authornote" and "authornotemark" commands
% used to denote shared contribution to the research.
\author{Ruotong Wang}
\authornotemark[1]
\email{ruotongw@cs.washington.edu}
\orcid{0000-0003-0964-6943}
\affiliation{%
  \institution{University of Washington}
  \city{Seattle}
  \state{Washington}
  \country{USA}
  \postcode{98195}
}

\author{Ruijia Cheng}
\authornote{This work is done during the author's internship at Microsoft Research.}
\email{rcheng6@uw.edu}
\orcid{0000-0002-2377-9550}
\affiliation{%
  \institution{University of Washington}
  \city{Seattle}
  \state{Washington}
  \country{USA}
  \postcode{98195}
}

\author{Denae Ford}
\email{denae@microsoft.com}
\orcid{0000-0003-0654-4335}
\affiliation{%
  \institution{Microsoft Research}
  \city{Redmond}
  \state{Washington}
  \country{USA}
  \postcode{98052}
}

\author{Thomas Zimmermann}
\email{tzimmer@microsoft.com}
\orcid{0000-0003-4905-1469}
\affiliation{%
  \institution{Microsoft Research}
  \city{Redmond}
  \state{Washington}
  \country{USA}
  \postcode{98052}
}

%
% By default, the full list of authors will be used in the page
% headers. Often, this list is too long, and will overlap
% other information printed in the page headers. This command allows
% the author to define a more concise list
% of authors' names for this purpose.
\renewcommand{\shortauthors}{Wang, et al.}

%%
%% The abstract is a short summary of the work to be presented in the
%% article.
\begin{abstract}
 Trust is a crucial factor for the adoption and responsible usage of generative AI tools in complex tasks such as software engineering. However, we have a limited understanding of how software developers evaluate the trustworthiness of AI-powered code generation tools in real-world settings. To address this gap, we conducted Study 1, an interview study with 17 developers who use AI-powered code generation tools in professional or personal settings. We found that developers' trust is rooted in the AI tool's perceived ability, integrity, and benevolence, and is situational, varying according to the context of usage. Existing AI code generation tools lack the affordances for developers to efficiently and effectively evaluate the trustworthiness of AI-powered code generation tools. To explore designs that can augment the existing interface of AI-powered code generation tools, we explored three sets of design concepts (suggestion quality indicators, usage stats, and control mechanisms) that derived from Study 1 findings. In Study 2, a design probe study with 12 developers, we investigated the potential of these design concepts to help developers make effective trust judgments. We discuss the implication of our findings on the design of AI-powered code generation tools and future research on trust in AI.

\end{abstract}
% \maketitle
% 

%%
%% The code below is generated by the tool at http://dl.acm.org/ccs.cfm.
%% Please copy and paste the code instead of the example below.
%%


%%
%% Keywords. The author(s) should pick words that accurately describe
%% the work being presented. Separate the keywords with commas.

%% A "teaser" image appears between the author and affiliation
%% information and the body of the document, and typically spans the
%% page.
% \begin{teaserfigure}
%   \includegraphics[width=\textwidth]{sampleteaser}
%   \caption{Seattle Mariners at Spring Training, 2010.}
%   \Description{Enjoying the baseball game from the third-base
%   seats. Ichiro Suzuki preparing to bat.}
%   \label{fig:teaser}
% \end{teaserfigure}

\def\authnote{1}
\newcommand{\fixme}[1]{\ifnum\authnote=1{\textcolor{red}{[FIXME: #1]}}\fi}

%%
%% This command processes the author and affiliation and title
%% information and builds the first part of the formatted document.
\maketitle

\section{Introduction}
With the rapid development of generative AI in recent years, it's increasingly used to support various human tasks in multiple domains, including complex information work such as software engineering. In software engineering, AI-powered code generation tools such as \gh Copilot~\cite{GitHubCopilotYour} and Tabnine~\cite{AIAssistantSoftware} have quickly gained popularity in programmer communities~\cite{Dohmke2022, liang2024large}, enabling a new way of programming assistance~\cite{sarkarWhatItProgram2022, barkeGroundedCopilotHow2022}. AI code generation tools can generate multiple lines of code in real-time based on a prompt within an Integrated Development Environment (IDE)~\cite{sarkarWhatItProgram2022}. 

While researchers and software developers are excited about AI-powered code generation tools, these tools also introduce new design challenges in creating responsible and reliable user experiences. One significant challenge involves helping users evaluate the trustworthiness of AI tools. 
Software developers' trust in programming support tools has long been studied as a crucial design requirement for such tools, as it serves as a key prerequisite for the safety of resulting software products~\cite{lipner2004,hasselbring2006, Zakir2014}. Without proper support, developers can find it challenging to form accurate mental models of what AI tools can do or not~\cite{sarkarWhatItProgram2022} or determine the quality of specific AI suggestions~\cite{vaithilingamExpectationVsExperience2022, barkeGroundedCopilotHow2022, pearceAsleepKeyboardAssessing2022}; thus becoming vulnerable to over- or under-trusting the AI~\cite{Zakir2014,murphyhill2021tse}. 

Existing research on trust in AI shows that the trustworthiness of technology is not inherent in an AI system but is based on how users interpret the information \textit{communicated} via the systems' interfaces and interactions~\cite{liaoDesigningResponsibleTrust2022}, and it can shift by context (e.g., task difficulty)~\cite{zhangShiftingTrustExamining2022, kimHumansAIContext2023}. Yet, while emerging work has begun investigating the general usability of generative AI assistant tools in software engineering or broader domains~\cite{vaithilingamExpectationVsExperience2022, barkeGroundedCopilotHow2022, sarkarWhatItProgram2022, zieglerProductivityAssessmentNeural2022, birdTakingFlightCopilot2023, weiszPerfectionNotRequired2021}, we still know little about how their interfaces should be designed to \textit{communicate} appropriate levels of trustworthiness and help developers form calibrated trust attitudes in AI-powered code generation tools. 

In this paper, we present results from a two-stage qualitative study. We started by getting an empirical understanding of developers' notions of trust in the particular context of using AI code-generation tools. In Study 1, we conducted interviews with 17 developers who have various levels of experience in using AI-powered code generation tools in real-life scenarios. We analyzed the results from Study 1 to answer the questions of \textit{what factors contribute to developers' trust attitudes in AI-powered code generation tools} (RQ1) and \textit{what challenges do developers face in evaluating the trustworthiness of AI tools} (RQ2). We found that developers evaluate the trustworthiness of an AI tool based on its perceived practical benefits, alignment
with their short- and long-term goals, and process integrity when generating outputs. Moreover, developers continuously reassess these factors in specific contexts based on situational factors such as stakes or complexity of tasks, forming situational trust attitudes. We also found that the lack of trust affordances in existing AI tools could result in inefficient and biased evaluation of AI's trustworthiness. To explore solutions to these challenges, we explored \textit{how to augment existing system interface to support effective and efficient evaluation of AI's trustworthiness} (RQ3) in Study 2. Specifically, we collected feedback on three groups of visual design concepts in design probe sessions with 12 additional developers. We found that design concepts, including quality indicators of AI suggestions, usage statistics dashboards, and control mechanisms to communicate user intention, show promise in scaffolding developers' trust judgments.

Our studies make the following contributions: 
(1) Building on prior literature that shows trust is rooted in the interplay between system characteristics and contexts ~\cite{mayerIntegrativeModelOrganizational1995a} and the call for empirical understanding of trust in specific application areas~\cite{kimHumansAIContext2023}, we provide a nuanced description of developers' notion of trust in generative AI tools in the context of programming, based on in-depth empirical data collected from interviews with developers who have real-world experience using AI powered code generation tools; (2) Furthering the growing literature on users' experiences with AI-powered code generation tools, we show the lack of ways to communicate users' intentions and lack of signals to validate AI output which are often characterized as usability challenges could, in fact, pose challenges for users to evaluate the trustworthiness of AI tools; (3) We contribute three groups of user-evaluated design implications, coupled with visual examples, to help designers take trust into consideration when designing AI-powered code generation tools.

\section{Related Work}

\subsection{Trust in AI}
Trust is considered a key factor affecting user interaction with AI~\cite{eu2019building, dasOpportunitiesChallengesExplainable2020, liaoDesigningResponsibleTrust2022}. The lack of trust can prevent users from adopting AI tools in their workflow, even when the system's performance is superior~\cite{boubinQuantifyingComplianceReliance2017a, oconnorQuestionTrustCan2019}. On the other hand, blind trust in AI, especially in high-stake tasks such as software engineering, can result in overlooking mistakes or risks produced by AI~\cite{pearceAsleepKeyboardAssessing2022, perryUsersWriteMore2022}. 

Trust in AI is defined as the user's attitude that ``an agent will help achieve an individual's goals in a situation characterized by uncertainty and vulnerability''~\cite{liaoDesigningResponsibleTrust2022, leeTrustAutomationDesigning2004, vereschakHowEvaluateTrust2021}, and therefore is particularly important when users engage in high-stake scenarios where the mistakes could have significant repercussions~\cite{jacoviFormalizingTrustArtificial2021a}.
% As a result, users' trust in AI is often studied in high-stake scenarios such as loan application~\cite{chen2018interpretable} and medical diagnosis~\cite{yangHarnessingBiomedicalLiterature2023}.
A review paper highlights that trust in AI is subjective and should be studied as an \textit{attitude}~\cite{vereschakHowEvaluateTrust2021}, distinguishing from reliance or compliance, which are often studied as a \textit{behavior}~\cite{tolmeijerCapableAmoralComparing2022}. Indeed, Mayer et al. characterized trust as ``an affective construct that can vary depending on the context and experiences of a person rather than simply being a rational or an objective reality.'' in their seminal paper on organizational trust~\cite{mayerIntegrativeModelOrganizational1995a}. In the context of AI, empirical evidence has also shown that users' trust is affected by contextual factors such as institution investment, other users' endorsement, and riskiness of task~\cite{widderTrustCollaborativeAutomation2021,leeTrustAutomationDesigning2004, zhangEffectConfidenceExplanation2020a}. 

Besides contextual factors, prior research identified that three system properties, including the system's ability, benevolence, and integrity, shape users' trust~\cite{kimHumansAIContext2023, mayerIntegrativeModelOrganizational1995a}. More recently, Liao et al. highlighted the importance of \emph{interface and interaction design} in mediating users' trust in AI~\cite{liaoDesigningResponsibleTrust2022}. Specifically, Liao et al. introduced the notion of \textit{trust affordances}, which are visual cues in the interface that indicate the system's trustworthiness. Users make trust judgments based on these trust affordances. Therefore, user interfaces and interactions play an important role in \emph{communicating} the internal trustworthy characteristics of AI to users. Following this understanding of trust, there has been a call for AI systems that can communicate an appropriate level of trustworthiness through their design, supporting users in building calibrated trust that aligns with the system's actual trustworthiness~\cite{yangHarnessingBiomedicalLiterature2023, jacoviFormalizingTrustArtificial2021a, bucincaProxyTasksSubjective2020}. 

Many prior HCI works have empirically investigated the effectiveness of various interface augmentations to support users in evaluating and calibrating trust. One common approach is to explain AI predictions and decisions using confidence score~\cite{zhangEffectConfidenceExplanation2020a, weiszPerfectionNotRequired2021} or visual explanations~\cite{yangHowVisualExplanations2020}, which could give users means and metrics to assess the performance of AI and make informed trust judgments. A related approach is to support the interpretability of model mechanisms~\cite{sunInvestigatingExplainabilityGenerative2022a, liaoQuestioningAIInforming2020, mishraCrowdsourcingEvaluatingConceptdriven2021}, increasing the predictability~\cite{daronnat2021inferring, drozdalTrustAutoMLExploring2020} of AI behavior. However, the effectiveness of these transparency features is not persistent across studies. For example, Agarwal et al. found that model confidence scores could mislead users' perception of the quality of model output~\cite{agarwalQualityEstimationInterpretability2021}. Another approach is to provide users with ways to control AI behavior. For example, research has shown that allowing users to co-create music with AI-powered tools~\cite{louieNoviceAIMusicCoCreation2020} or collaborate on writing tasks with AI models~\cite{leeCoAuthorDesigningHumanAI2022} can foster a sense of control and ownership, leading to higher trust in the system. Lastly, it has been shown that cognitive forcing functions such as delay showing AI's output could encourage users to engage with AI output analytically and reduce over-trust in AI~\cite{bucincaTrustThinkCognitive2021}. 

Despite the plethora of research on trust in AI, most centers on deterministic AI tools for classification or prediction tasks~\cite{vereschakHowEvaluateTrust2021}. These studies highlight the opportunity to support users in evaluating the trustworthiness of generative AI systems and forming calibrated trust attitudes using interfaces and interactions. However, how existing insights translate to generative AI tools, especially in software engineering contexts, remains an open question. Characteristics such as the richer and more complex input and output space~\cite{sunInvestigatingExplainabilityGenerative2022a} and more flexible roles in human-AI collaboration~\cite{guzdialFriendCollaboratorStudent2019} distinguish generative AI tools from the deterministic AI tools that are widely studied, but also introduces new challenges and opportunities in designing for users' trust. 
% To bridge this knowledge gap, we conducted an interview study with developers who have experience using AI-powered code generation tools in real life to understand their needs and challenges in making trust judgments and a design probe study to explore trust affordances that can support developers in evaluating the trustworthiness of AI code generation tools.  

\subsection{Generative AI in software engineering: AI-powered code generation tools}

The recent development of generative AI models unleashes new possibilities for AI tools to support complex human tasks~\cite{bommasaniOpportunitiesRisksFoundation2021}, including software engineering~\cite{barkeGroundedCopilotHow2022, zieglerProductivityAssessmentNeural2022, birdTakingFlightCopilot2023, weiszBetterTogetherEvaluation2022}. 
% Generative AI models produce textual or image artifacts in response to human prompts, resulting in the potential of supporting more creative human tasks such as writing~\cite{geroMetaphoriaAlgorithmicCompanion2019, cheng-etal-2022-mapping, clarkCreativeWritingMachine2018}, drawing~\cite{ohLeadYouHelp2018, fanCollabdrawEnvironmentCollaborative2019}, game design~\cite{guzdialFriendCollaboratorStudent2019}, music creation~\cite{louieNoviceAIMusicCoCreation2020} and software engineering~\cite{barkeGroundedCopilotHow2022, zieglerProductivityAssessmentNeural2022, birdTakingFlightCopilot2023, weiszBetterTogetherEvaluation2022}. 
% Building on prior work exploring the emerging interaction paradigms between users and generative AI models that are mostly in creative tasks, we study user interactions with generative AI tools in the context of software engineering. 
Software engineering is a type of complex and high-demanding information work that often involves high cognitive load and stress~\cite{sarkarWhatItProgram2022, ernst2022ieeesoftware, Helgesson2019, Goncales2019}, and therefore demands high-quality support. As a means to provide support for this complex work, commercial AI-powered code generation tools such as \gh Copilot~\cite{GitHubCopilotYour} and Tabnine~\cite{AIAssistantSoftware} have emerged and become a novel service to expert and novice code creators alike.
%that have been integrated into developers' workflows~\cite{barkeGroundedCopilotHow2022, zieglerProductivityAssessmentNeural2022, birdTakingFlightCopilot2023}. 
These tools provide AI services powered by large language models trained on code data~\cite{GitHubCopilotYour, sobania2022comprehensive} and suggest code based on user prompts and project context~\cite{brown2020language}. For instance, powered by the OpenAI Codex model~\cite{OpenAICodex2021}, Copilot is an extension in code editors that can generate code suggestions as ghost text at the user's cursor location. When using AI code generation tools, users can write comments in natural language and prompt the AI to generate code that they can accept, reject, make edits, and choose from various candidates. The AI can also complete users' in-progress code within a single line or by completing the function. Compared to traditional code completion tools based on defined rules and documentation, AI-powered code generation tools produce longer and more contextually relevant code snippets by synthesizing new code that might not exist in any code base~\cite{birdTakingFlightCopilot2023}. 
% sdes are complex knowledge workers, high skills, high impact,
As AI code generation tools introduce a brand new interaction paradigm between developers and AI~\cite{mozannarReadingLinesModeling2023, xuInIDECodeGeneration2022}, early empirical investigations show that developers struggle to adapt to the new interaction pattern, often having an incomplete mental model of what Copilot can do or not~\cite{sarkarWhatItProgram2022} and finding it challenging to review and evaluate the quality of AI-generated code~\cite{vaithilingamExpectationVsExperience2022, barkeGroundedCopilotHow2022, birdTakingFlightCopilot2023, perryUsersWriteMore2022}. These known usability challenges motivate us to understand how developers evaluate the trustworthiness of AI-powered code-generation tools.  

While user trust has long been considered crucial in the design of traditional programming support tools, such as compilers and version control systems~\cite{shundan2014cscw,witschey2015} to ensure software safety~\cite{lipner2004,hasselbring2006}, developers' trust in AI code generation tools needs even more nuanced and careful consideration
~\cite{pearceAsleepKeyboardAssessing2022, Dakhel2022, birdTakingFlightCopilot2023} due to the uncertainty introduced by generative AI. For example, since the mechanism of generative AI is more opaque and the outputs are more difficult to anticipate than traditional developer support tools, developers must establish an appropriate level of trust with these AI tools and be cautious about the potential risks~\cite{birdTakingFlightCopilot2023}. In particular, Widder et al. conducted an ethnography study in 2021 with developers who use deterministic code generation tools and uncovered 16 factors that affect their trust in the tool~\cite{widderTrustCollaborativeAutomation2021}. Given that trust is deeply embedded in contextual factors, the factors identified in prior empirical studies might change for AI-powered code-generation tools as system properties and the social and organizational contexts of usage shift.

The pressing need to support developers in building and calibrating trust and the gap in previous literature motivated us to conduct retrospective interviews with developers who have experience using AI code generation tools in real-life scenarios (Study 1). In our research setting, developers face real consequences if AI produces undesirable outcomes, which could help us uncover the interplay of trust and contexts. Building on the interview findings and literature on the design of trust affordances, such as transparency and users' control, we further conducted a design probe study to understand the design space of trust affordances that can support developers in evaluating the trustworthiness of AI-powered code generation tools (Study 2). 

\section{Study 1: How do developers evaluate the trustworthiness of AI tools?}
To understand what contributes to developers' trust attitudes in AI code generation tools (RQ1), as well as their challenges in making trust judgments (RQ2), we conducted retrospective interviews with 17 developers who use AI-powered
code generation tools in real-life professional or personal settings. 

\subsection{Methods: Retrospective Interview Study}
\subsubsection{Study Procedure: collect critical incident + retrospective interview}
To capture the interplay between trust and specific contexts of usage, we adopt a method of critical incident sampling and retrospective interviews. A similar approach has been applied to study patients' trust during medical visits~\cite{yanez-gallardoCriticalIncidentsTrust2012, wendtTrustConfirmationGynecologic2004} and interpersonal trust in business negotiations~\cite{munscher201116}. A week before the scheduled interview, we contacted participants via instant message and asked them to prepare for the interview by collecting their significant moments when using AI-powered code generation tools during the following week---i.e., moments where they were either particularly satisfied, disappointed, or surprised. Participants were asked to share the descriptions and screenshots of those moments with us and were reminded regularly throughout the week. These records of significant moments helped participants recall the nuances of their experience in the interviews, allowing us to understand their trust in AI tools in realistic contexts of use. 
During the 60-minute retrospective interview sessions, we asked participants about their general experience with AI tools and then asked them to walk through the significant moments they collected during the prior week. We specifically probed for factors that affected their trust attitudes in AI. The interviews were conducted from July to August 2022, and the study procedures received approval from the Institution Review Board.

\subsubsection{Participants}
We recruited 17 software engineers with diverse programming and AI tools experience. Participants were recruited from different organizations of a large technology company through messages shared in group chats and emails distributed to developers chosen randomly from a directory. We stopped recruiting after hearing repeating themes in the interviews. Our final sample consists of 15 male and two female participants, aged between 18 and 54 years, with varying degrees of work experience and seniority. Participants had programming experience ranging from 2-25 years. They reported working on different areas of development (e.g., front-end, back-end, data science) and were involved in various types of development tasks (e.g., modifying existing features, writing new features, writing tests, refactoring). All participants had experience using Github Copilot, with various frequencies (9 daily, 3 weekly, 3 monthly, 2 recently started) and experience using it in professional and personal settings. Two also had experience with Tabnine. Detailed profiles of our participants are included in the Appendix (Table~\ref{tab:study1-participant}).

\subsubsection{Data Analysis}
All interviews were video and audio recorded and later transcribed. Our analysis of the interview data followed the procedure of inductive thematic analysis~\cite{braunReflectingReflexiveThematic2019}. The first two authors took detailed field notes and frequently discussed the emerging themes with the research team during data collection. Based on the field notes and discussions with the research team, the first author developed an initial codebook, applied it to the interview data, and noted places where codes could be merged or refined. The research team then collectively refined and grouped the codes via discussion, deriving a final code book, which is then re-applied to the data. The final codebook consists of 39 codes that focus on factors affecting developers' general trust in AI tools, their process of evaluating specific AI suggestions, and their challenges in building trust in AI tools. Example codes include ``\textit{trust varied by situations}'', ``\textit{initial expectation affects trust building}'' or ``\textit{trial and error to build trust}''.

\subsection{Factors that contribute to developers' trust attitudes in AI tools (RQ1)}
Aligning with prior literature that indicates systems' \emph{ability, integrity,} and \emph{benevolence} as important factors that contribute to users' trust attitudes~\cite{mayerIntegrativeModelOrganizational1995a, liaoDesigningResponsibleTrust2022}, we observed that developers trust a given AI-powered code generation tool when they perceive practical benefits (\emph{ability}), alignment
with their goals (\emph{benevolence}), and trustworthy processes (\emph{integrity}). We also observed that \emph{situational factors}, such as stakes of the use scenario and the complexity of the programming task, mediate developers' trust attitudes.  

\subsubsection{Ability: AI tools' practical benefits}~\label{sec:study1_assess_ability}
The ability of an AI tool is defined as its competence or performance~\cite{leeTrustAutomationDesigning2004, mayerIntegrativeModelOrganizational1995a, liaoDesigningResponsibleTrust2022}. In the context of AI code generation tools, we observed that developers commonly assess the ability of an AI tool based on its \textbf{practical benefits to their work}, often related to time saved or lines of code contributed. Instead of expecting AI to provide perfect solutions, developers value the ease of building upon AI's outputs. Even when recognizing that AI's suggestions ``\textit{may not be able to compile or run correctly}'', P16 still trusts the tool: ``\textit{because I can always go back and modify it a little bit, tune it maybe, and get it to output what I want.}'' P10 values AI's utility in ``\textit{lay(ing) the foundation very well.}'' At the same time, P13 pointed out the potential for trust erosion if the AI's suggestion requires extra time to verify and correct: ``\textit{If Copilot ever slows down \dots I would consider not using Copilot anymore.}''

% While most developers agree that demonstrating AI's utility helps them build trust, the definition of utility is subjective and can vary significantly. Some developers see the utility of AI tools beyond just saving time and increasing productivity but also in enhancing their thinking processes. For example, P14 trusts Copilot because it serves ``\textit{as a tool to slow me down, provide a different perspective on what I'm implementing...forcing me to confront my own thought process.}'' Similarly, For example, P15 trusts Copilot because it can scaffold learning: ``\textit{it would prompt me for ways to do it that maybe I wouldn't have thought of myself},'' which helped them pick up a new programming language more quickly.

\subsubsection{Benevolence: alignment of goals between AI and developer}~\label{sec:study1_assess_benevolence}
Benevolence refers to the alignment of the AI tool's goals and users' goals~\cite{mayerIntegrativeModelOrganizational1995a, liaoDesigningResponsibleTrust2022}. When it comes to AI code generation tools, benevolence is the perception that the AI tool is designed with developers' best interests in mind, supporting not just immediate task completion but also their long-term goals, such as learning and career growth. Trust arises when developers are convinced that the AI tool \textbf{respects their personal preferences, learning goals, and career aspirations}. However, we observed many instances of distrust due to the mismatch between what developers expect from the AI and the tool’s actual behavior, leaving the impression of AI being aggressive and obtrusive. Regarding immediate task completion, 
P13 often found AI's suggestions to create unnecessary ``\textit{visual clutter}'' on the screen when they already knew what they wanted to write. P7 felt that they had to ``\textit{fight with Copilot}'' to let unwanted suggestions go away. 
Developers also worry that using AI tools would hinder their personal and career growth in the long run (i.e., limiting learning opportunities or eventually replacing their jobs).
For example, after accepting high-quality suggestions from AI for a while, P8 started to worry about losing their ``\textit{programming muscles}.'' As P8 said, Copilot ``\textit{want to sit in my seat...It started as a co-pilot, but now it's the pilot and I'm becoming the co-pilot.}'' P7 echoed the sentiment and worried that ``\textit{(AI tool is) robbing me from the opportunity to actually use my brain,''} preventing them from improving their own programming skills.

\subsubsection{Integrity: the model mechanisms}~\label{sec:study1_assess_integrity}
Integrity is defined as whether the operational process of AI is appropriate to achieve users’ goals (e.g., fair and secure when making decisions)~\cite{liaoDesigningResponsibleTrust2022, mehrotraIntegrityBasedExplanations2023}. Developers trust AI tools when they are informed about and agree with \textbf{the model mechanism}. As P17 puts it: ``\textit{knowing how it works gives me more trust because I think it's just whether I agree with your approach or not.}'' Developers specifically highlighted the need to understand AI tools' security and privacy implications. P5's trust in Copilot increased after reading that ``\textit{it's bound by all these privacy laws.}'' Others noted a lack of relevant information for them to understand AI tools' process integrity. 
% For instance, P7 sought information on code ownership, and P9 was concerned about their programming process being ``\textit{listened to}'' and wanted clarification on the AI's ``\textit{boundaries}''. Similarly, 
For example, P16 desired ``\textit{an end-to-end transparent diagram with what's exactly going on}'' so that they could know ``\textit{exactly what's being tested to make sure that this code is appropriately copyrighted}.''

\subsubsection{Situational factors: the stake and complexity of tasks}~\label{sec:study1_situational}
Developers' trust in AI tools is not an object translation of the tool's ability, benevolence, and integrity but a dynamic assessment of the system characteristics together with additional situational factors such as \textbf{the stakes of the scenario} or \textbf{the complexity of tasks}.
Developers are more reluctant to trust AI tools in high-stake and high-impact scenarios, such as on codebases that could ``\textit{impact millions of customers and millions of dollars potentially}''(P13). In those cases, they would only allow AI tools to play a ``\textit{suggestive}'' role (P10) instead of generating code that would go into production. In another example, P3 shared that they would trust Copilot when writing ``\textit{proof-of-concept type of the projects},'' but not in ``\textit{actual production setting}.''
The complexity of tasks also affects trust. While P2 trusts AI tools for smaller mundane tasks that are ``\textit{standard and common}'' and ``\textit{involves less logic to do},'' they don't expect AI tools to be useful for ``\textit{open-ended stuff.}''
Others also decide against using AI tools in situations with special requirements due to a lack of trust. For example, P6 does not trust Copilot to generate code that satisfies accessibility and responsiveness requirements. P2 does not use Copilot when they need to share the code with others because they don't trust it to generate code "\textit{in the most explainable way that other people would understand}.". 

\subsection{Challenges in evaluating the trustworthiness of AI tools (RQ2)}
While AI tools' ability, benevolence, integrity and situational inform developers' trust attitudes, our findings show that the design of current AI-powered code generation tools fails to adequately support developers to evaluate these factors, leading to inefficiency and bias in trust attitudes. We outline three key challenges in this section.   

\subsubsection{Biased trust attitudes due to lack of reliable source of information on AI ability}~\label{sec:study1_challenge_limited_exposure}
Given that the performance of generative AI varies greatly depending on the specific context and task, making informed trust judgments requires developers to have a clear understanding of AI tools' ability in different situations. However, we notice a lack of reliable sources of information for developers to understand the ability of AI tools. As a result, developers commonly rely on intuitions accumulated from first-hand experiences of evaluating AI outputs to determine AI tools' abilities in different situations. Developers like P13 form intuitions by observing AI performance in routine programming tasks: ``\textit{once you've seen it 10 times, I'm pretty sure Copilot will do this thing the 11th time.}'' Others like P6 ``\textit{played around}'' or intentionally experiment with AI to try to ``\textit{break it'}' when first started using Copilot, so that they can ``\textit{know where its limits are}'', which helped ``\textit{set my expectations on how to use it.}'' 

However, the sole reliance on developers' personal experience can be inefficient. P13 shared that: 
``\textit{you have to give it the benefit of the doubt for a while until it makes a little more sense to you as a tool...You just have to ignore those things until your expectation lines up with Copilot's capabilities.}'' It also leads to biased perceptions of the trustworthiness of AI tools. We observed that while positive experiences with AI tools lead to increased trust, negative experiences disproportionately impact developers' trust. A single misstep could instantly undermine their trust. For example, when P5 started using Copilot and found its multi-line suggestion to be unhelpful, they decided that they would ``\textit{not even read into the [multi-line] recommendations}.'' P5 further emphasized that: ``\textit{it takes three good recommendations to build trust versus one bad recommendation to lose trust}.''
The issue is exacerbated when developers bring expectations from traditional non-AI-based auto-completion tools such as IntelliSense, which pulls error-free code directly from the documentation. This creates unrealistic expectations for AI tools that generate more flexible suggestions that usually require reviews and edits and could lead to disappointment. P17 commented that ``\textit{It (AI tools) has to be better than IntelliSense to be worth using.}'' P14 also shared their frustration when observed that Copilot did not give them ``\textit{the right solution}'' or, in fact, the same solution as IntelliSense. 

% ``\textit{Trust really is dependent on the empirical evidences}'', as P9 concisely put. 
% P15 echoed that ``\textit{hands-on user experiences are what shine the most light on how much you can trust it}'' and explained: ``\textit{everyone's writing code in different circumstances... You can never know exactly how well it's going to perform in your unique situation until you try it.}''

% P7 also shared their experience: ``\textit{if it demonstrated over time that it was useful, then I'd maybe use it more and more.}'' 
% For P10, seeing examples of Copilot suggestions that failed to pick the correct method made them ``\textit{trust Copilot less in these scenarios.}'' 

% They wanted to see an AI tool \textit{demonstrating} its values in practice and to ``\textit{feel the benefits of it}'' (P17) in order to build up positive expectations.  

% Developers share the common desire to validate AI suggestions before accepting them. Developers often wanted to maintain the final say on AI suggestions to catch any problematic output in time: ``\textit{[I] would very carefully look at what it was suggesting before I hit enter tab because I had to be certain.}'' (P7) 

\subsubsection{Ineffective and inefficient evaluation of AI output}~\label{sec:study1_challenge_evaluate}
While evaluating each specific instance of AI suggestions forms the basis of developers' understanding of the AI tool's ability, we observed that developers often rely on inefficient manual methods due to inadequate support for evaluating the suggestion quality. The common strategies that developers use, such as ``\textit{logically going through the problem,}'' (P11) or ``\textit{validat[ing] by testing it,}'' (P13), can be time-consuming and ineffective. P9 shared that they
spent half an hour identifying a small error of an additional bracket in a long block of code suggestions that spanned multiple lines. Some developers turn to external tools such as refactoring tools or library documentation for assistance. However, frequently using these methods could disrupt their programming workflow. The process of constantly switching between writing and reviewing code was described to be ``\textit{mentally draining}'' (P7), ``\textit{derail my mind.}'' (P9) and eventually ``\textit{creates more work}'' (P8). For example, P1 once had to spend extra effort researching a method they were not familiar with to debug: ``\textit{I had to look back at documentation, and it was using fields that were deprecated or nonsense fields that just created on its own}.''

\subsubsection{Lack of mechanisms to align AI with developers' goals and preferences}~\label{sec:study1_challenge_prompting}
Aligned goals between developers and AI tools indicate the benevolence of the tool. However, the current interaction paradigms of AI code generation tools that mostly rely on including information in in-progress code for the AI to produce desired outputs make it challenging for developers to communicate their short-term and long-term goals and preferences to AI tools, not to mention signaling the benevolence of the tool. P1 and P8 find it difficult to tailor their prompts to guide AI output without sacrificing their programming flow. As a result, they chose not to trust AI suggestions because ``\textit{there's no reason to expect Copilot will read my mind and figure out what I want to do now.}'' (P1) P9 desired a ``\textit{sensei}'' version of Copilot that is more ``\textit{endearing}'' and would ``\textit{invest in you by suggesting what you could learn}'', but find no way to communicate their goal.
Another common challenge is signaling the desired timing of suggestions from AI tools. Many developers expressed frustration that they did not get enough suggestions when they desired AI's help; whereas other times they found AI suggestions to be intrusive, getting in the way of their programming flow. For example, P11 did not want ``\textit{Copilot to jump the gun and suggest before I finish fully defining the method}'' because it would likely lead to ``\textit{suggestions that are way off the mark.}''

\subsection{Summary of results}
Our findings in Study 1 reveal that developers tend to trust AI tools when they perceive practical benefits, alignment with their goals, and trustworthy processes. Furthermore, developers adjust their trust by considering additional situational factors such as task complexity and importance. However, the current AI tools do not provide enough support for the developers to assess AI tools' ability and benevolence in specific situations, resulting in inefficient and biased evaluation of the trustworthiness of AI tools. These findings motivate us to explore ways to improve the existing interface and interaction design of AI code generation tools to help developers more effectively calibrate their trust attitudes.

\section{Study 2: How to support developers to evaluate the trustworthiness of AI tools?}
We conducted a design probe study to further explore how to augment existing system interfaces to support effective and efficient evaluation of AI’s trustworthiness (RQ3). Building on findings in Study 1, we developed three groups of design concepts with visual representations and collected feedback from 12 developers. Notably, we do not aim to settle on or quantitatively evaluate the effectiveness of any specific design---rather, we used the designs as stimuli to elicit developers’ feedback and aim to explore the potential of these interface design concepts as trust affordances through nuanced qualitative exploration. A similar approach has been used to explore interface designs for AI-assisted decision-making systems in child welfare~\cite{kawakamiX201cWhyCare2022} and clinical diagnosis~\cite{yangHarnessingBiomedicalLiterature2023}. 

\subsection{Developing design concepts and visual stimuli}
% We identified three groups of design concepts, each of which addresses one of the challenges uncovered in Study 1. To create the visual representations of the three groups of design concepts, the team 
We first brainstormed design concepts that can address each of the challenges from Study 1. Some concepts were directly inspired by participants' interviews (e.g., control of timing, confidence score), while others were informed by literature (e.g., XAI features in~\cite{liaoDesigningResponsibleTrust2022}, uncertainty visualizer in~\cite{sunInvestigatingExplainabilityGenerative2022a}). Once the team settled on the three groups of concepts, the first author created the initial visuals, which were then iterated with additional feedback from the team and pilot participants. We intentionally kept the visuals low-fidelity since we wanted participants to focus on evaluating the \textit{high-level concepts} of the design instead of the usability of specific graphical or textual elements, following the suggestion in~\cite{10.5555/1526229}. All visual representations follow the design style of Copilot in Visual Studio Code since this combination was most commonly mentioned by participants in Study 1. In this subsection, we highlight the main features of each design concept and include the full visual representations in Appendix~\ref{fig:design_concepts_appendix}).

\subsubsection{Usage statistics dashboard to allow structured reflection on AI capability}

\begin{figure}[h]
    \centering
     \begin{subfigure}[b]{0.43\textwidth}
         \centering
         \includegraphics[width=\textwidth]{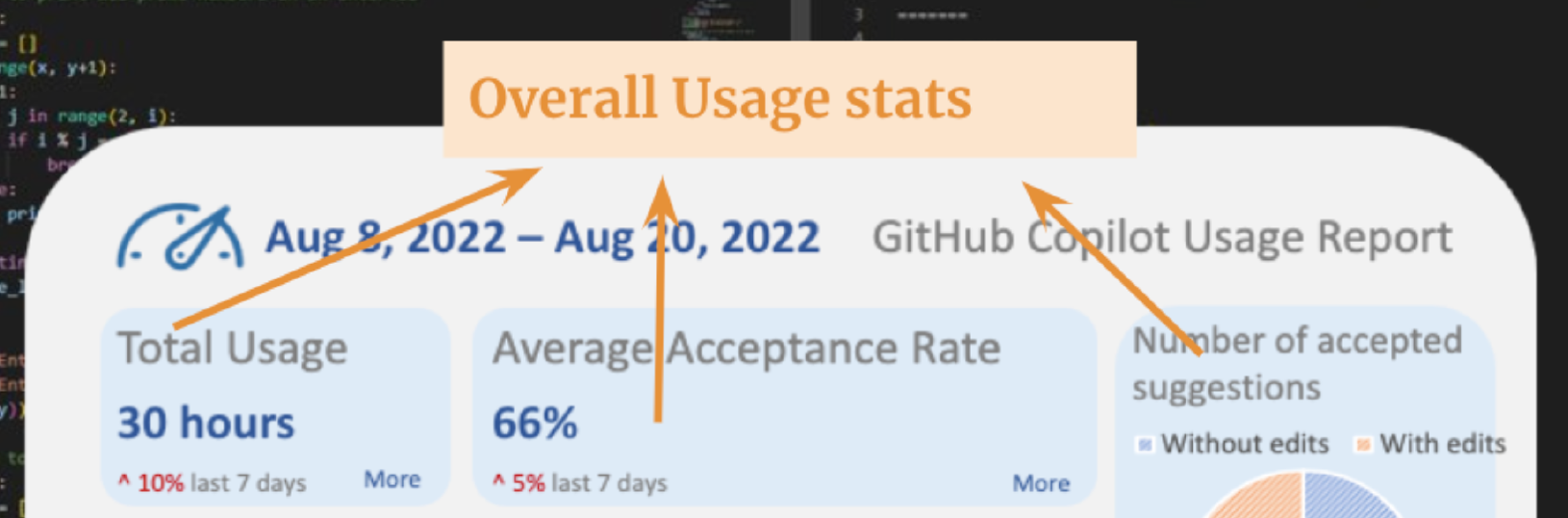}
         \caption{Overall usage stats}
         \Description{ }
         \label{fig:Overall usage stats}
     \end{subfigure}
     \hfill
     \begin{subfigure}[b]{0.45\textwidth}
         \centering
         \includegraphics[width=\textwidth]{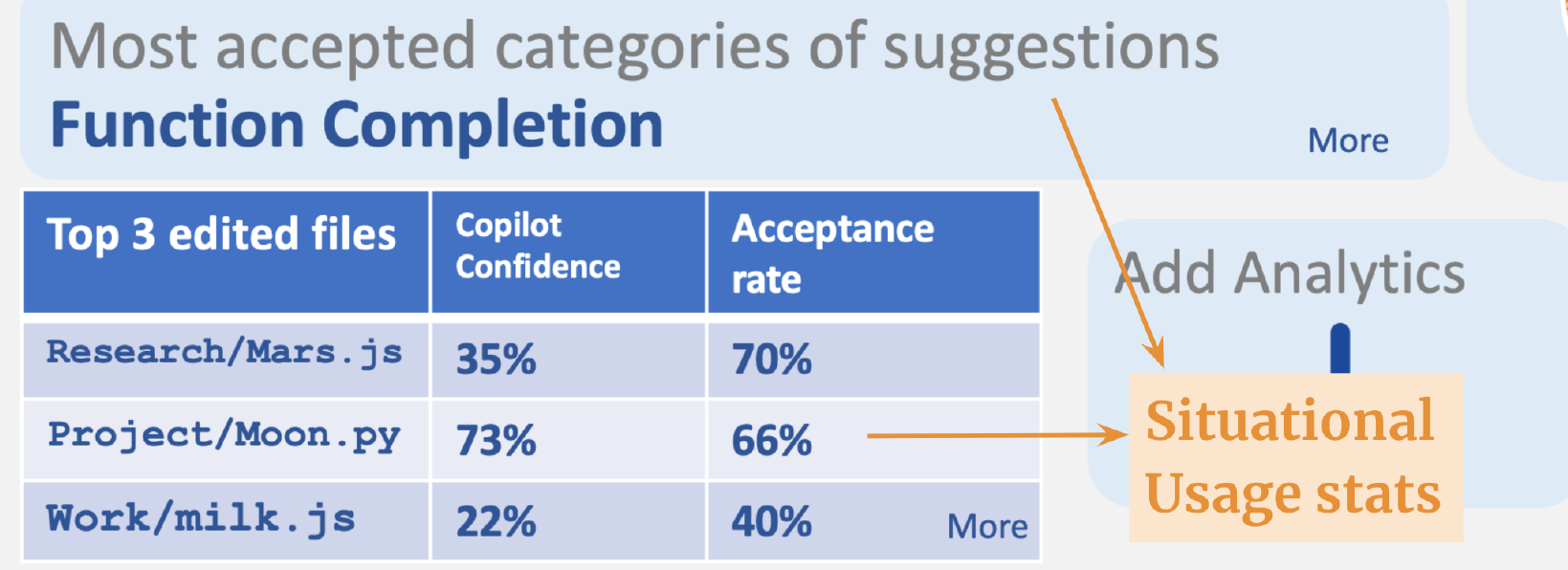}
         \caption{Situational usage stats}
         \Description{}
         \label{fig:Situational usage stats}
     \end{subfigure}
     \caption{A usage statistics dashboard that displays personalized usage statistics to a user. Both (a) overall usage stats and (b) situational usage stats are shown in a pop-up dashboard in IDE. }
     \label{fig:usage_stats}
\end{figure}

Study 1 reveals that developers' sole reliance on intuitions accumulated through personal experiences can lead to biased assessments of AI tool's trustworthiness in different situations (§~\ref{sec:study1_challenge_limited_exposure}). This points to a need for a more \textbf{structured approach to explicitly communicate AI tool's strengths, limitations, and applicability in specific contexts} to help developers understand and reflect on their trust attitude. Specifically, we designed a dashboard that displays personalized usage statistics to developers, with comparisons with AI tools' objective performance metrics in specific situations. The dashboard appears as a pop-up in the IDE after a user has used the AI tool for a certain period of time. The dashboard contains users' overall usage stats (Figure ~\ref{fig:Overall usage stats}) and usage stats broken down by files (Figure ~\ref{fig:Situational usage stats}). The overall usage statistics include data such as total hours of usage and average acceptance rate, which help developers reflect on their interaction with the AI tool. The situational usage statistics include data such as the most accepted categories of suggestions, which enable developers to calibrate their trust according to different contexts. The comparisons between users' acceptance rates and AI tools' confidence in different contexts serve as a reality check against developers' expectations. Users can access the dashboard via a button whenever they want to see it, allowing developers to dynamically recalibrate their trust based on ongoing usage. 

\subsubsection{Quality indicators to support efficient in-context evaluation of AI suggestions}
\begin{figure*}[h]
      \centering
     \begin{subfigure}[b]{0.60\textwidth}
         \centering
         \includegraphics[width=\textwidth]{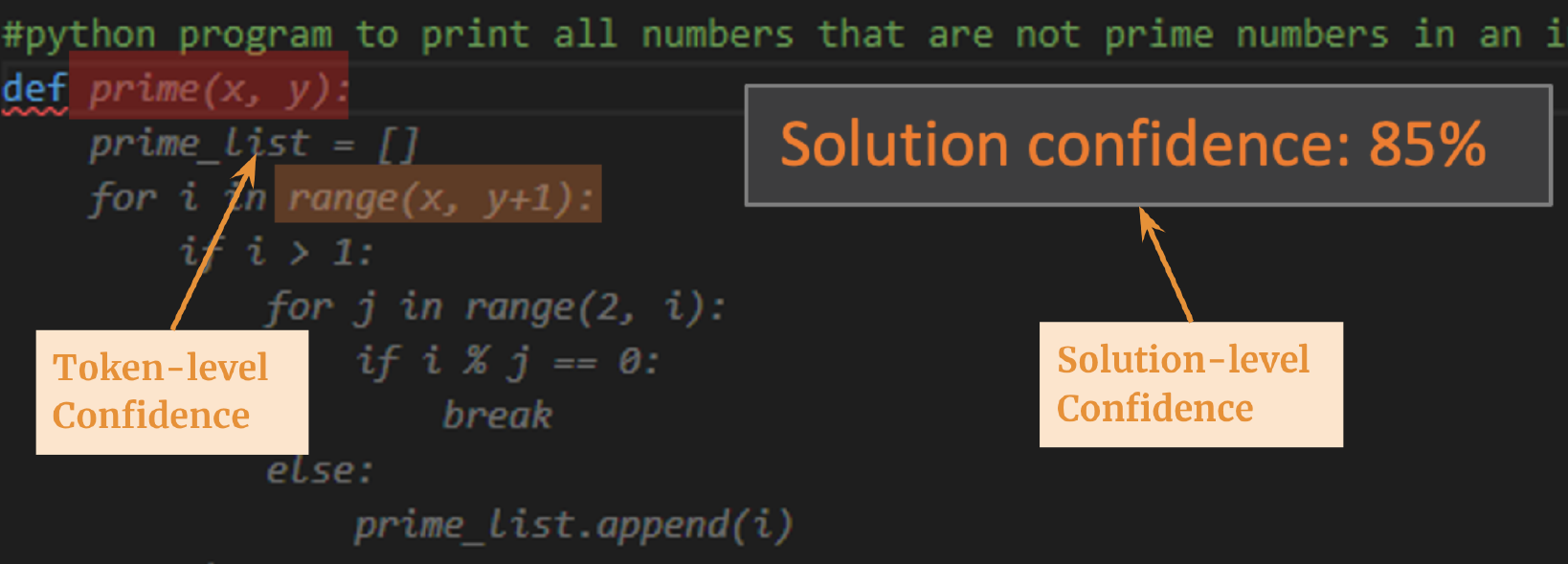}
         \caption{Solution-level and token-level confidence explanations}
         \Description{Solution-level and token-level confidence explanations}
         \label{fig:local_transparency}
     \end{subfigure}
     \hfill
     \begin{subfigure}[b]{0.35\textwidth}
         \centering
         \includegraphics[width=\textwidth]{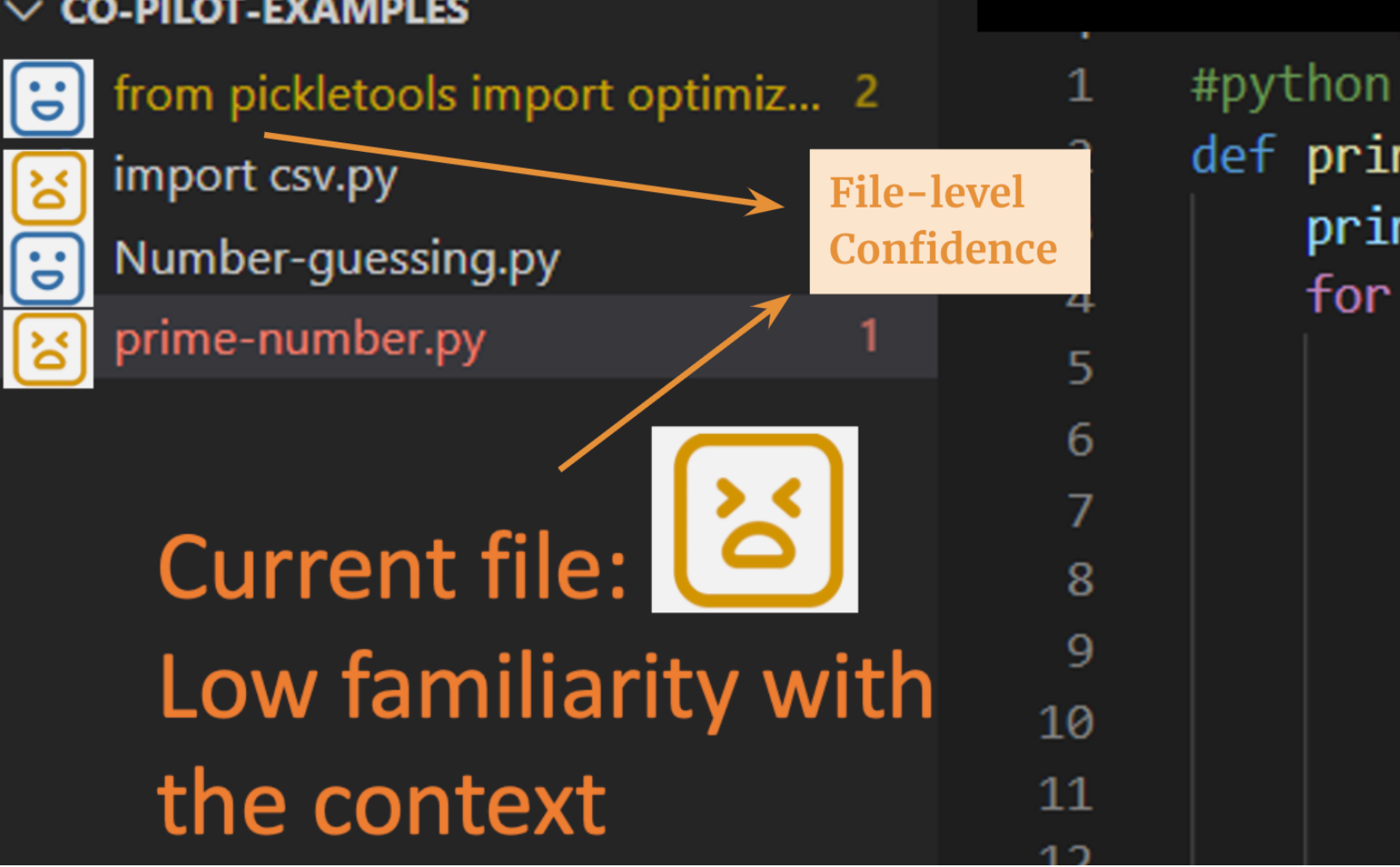}
         \caption{File-level familiarity explanation}
         \Description{File-level familiarity explanation}
         \label{fig:file_transparency}
     \end{subfigure}
     \centering
     \caption{Quality indicators to support users better evaluate each AI suggestion.}
     \label{fig:transparency_overall}
\end{figure*}

Evaluating each instance of AI suggestions helps developers build up their understanding of AI's abilities and enables developers to integrate AI output into their workflow (§~\ref{sec:study1_assess_ability}). However, the lack of support for the evaluation process forced developers to rely on manual methods or external tools (i.e., documentation), which are often time-consuming, ineffective, and disrupt developers' workflow (§~\ref{sec:study1_challenge_evaluate}). Thus, we created design concepts to \textbf{provide in-context support that enhances the evaluation process without disrupting the workflow}. Concretely, we explored three ways to provide transparency into the AI model's confidence in the output as non-disruptive ways to help developers make quick and accurate assessments of the quality of suggestions. The \textit{Solution-level confidence explanation} (Figure~\ref{fig:local_transparency}) indicates the model's aggregated confidence of the solution in the editing window, helping developers to quickly decide whether to build upon the AI's suggestion or discard it. If developers decide to scrutinize the suggestion closely, the \textit{Token level confidence/uncertainty explanation} (Figure~\ref{fig:local_transparency}) highlights specific tokens in the solution where the model has low confidence, helping developers to identify potential problems in the suggestion. Finally, the \textit{File-level familiarity explanation} (Figure~\ref{fig:file_transparency}) communicates the model's familiarity and alignment with the specific context in the file. For example, if the model has not seen input in the specific programming language or is using a particular library, the familiarity indicator might turn yellow or red to indicate that the model is unfamiliar with the context provided in the file. 

\subsubsection{Control mechanisms at the onset and during programming session to help align developers and AI's goals} 
\begin{figure}[h]
    \centering
     \begin{subfigure}[b]{0.46\textwidth}
         \centering
         \includegraphics[width=\textwidth]{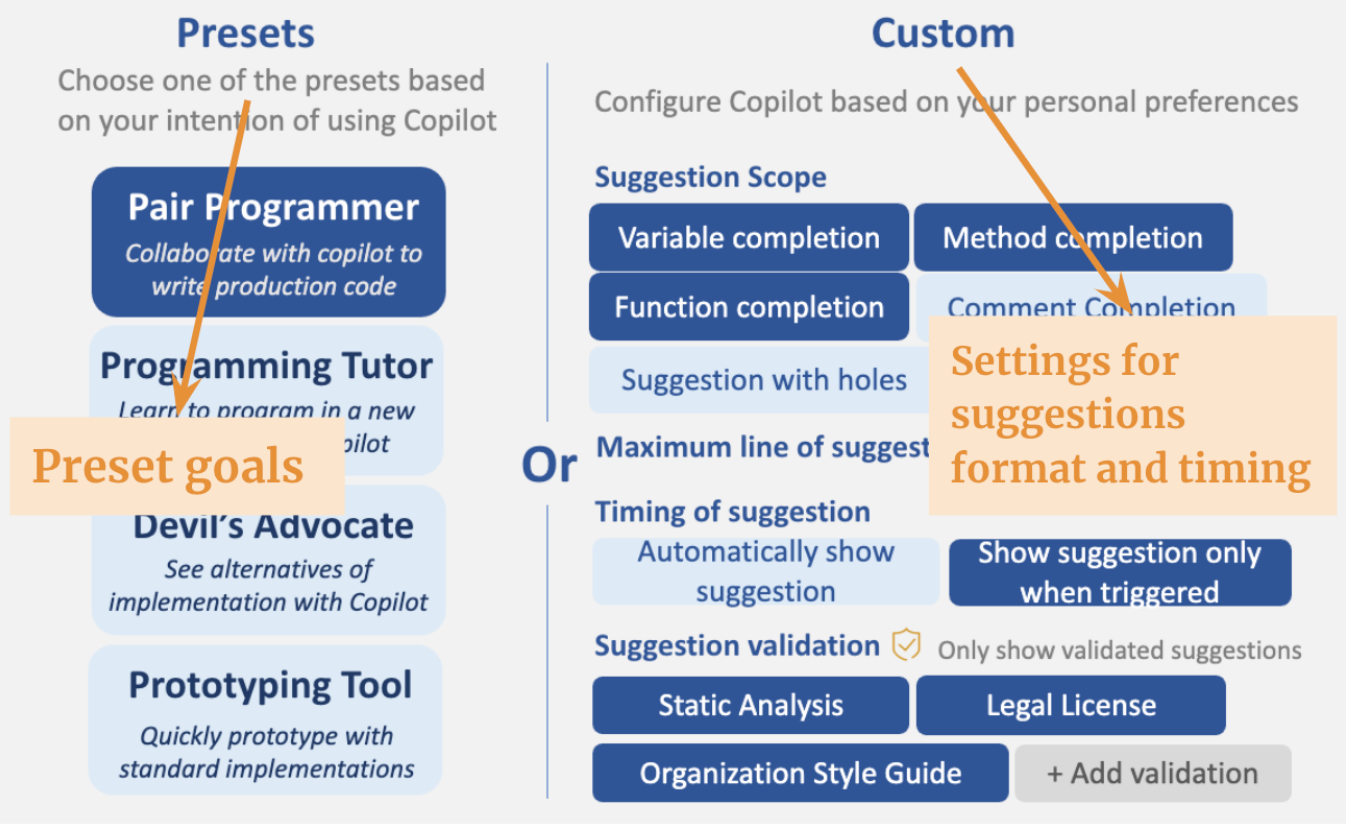}
         \caption{Control panel at the project initialization}
         \Description{ }
         \label{fig:control}
     \end{subfigure}
     \hfill
     \begin{subfigure}[b]{0.44\textwidth}
         \centering
         \includegraphics[width=\textwidth]{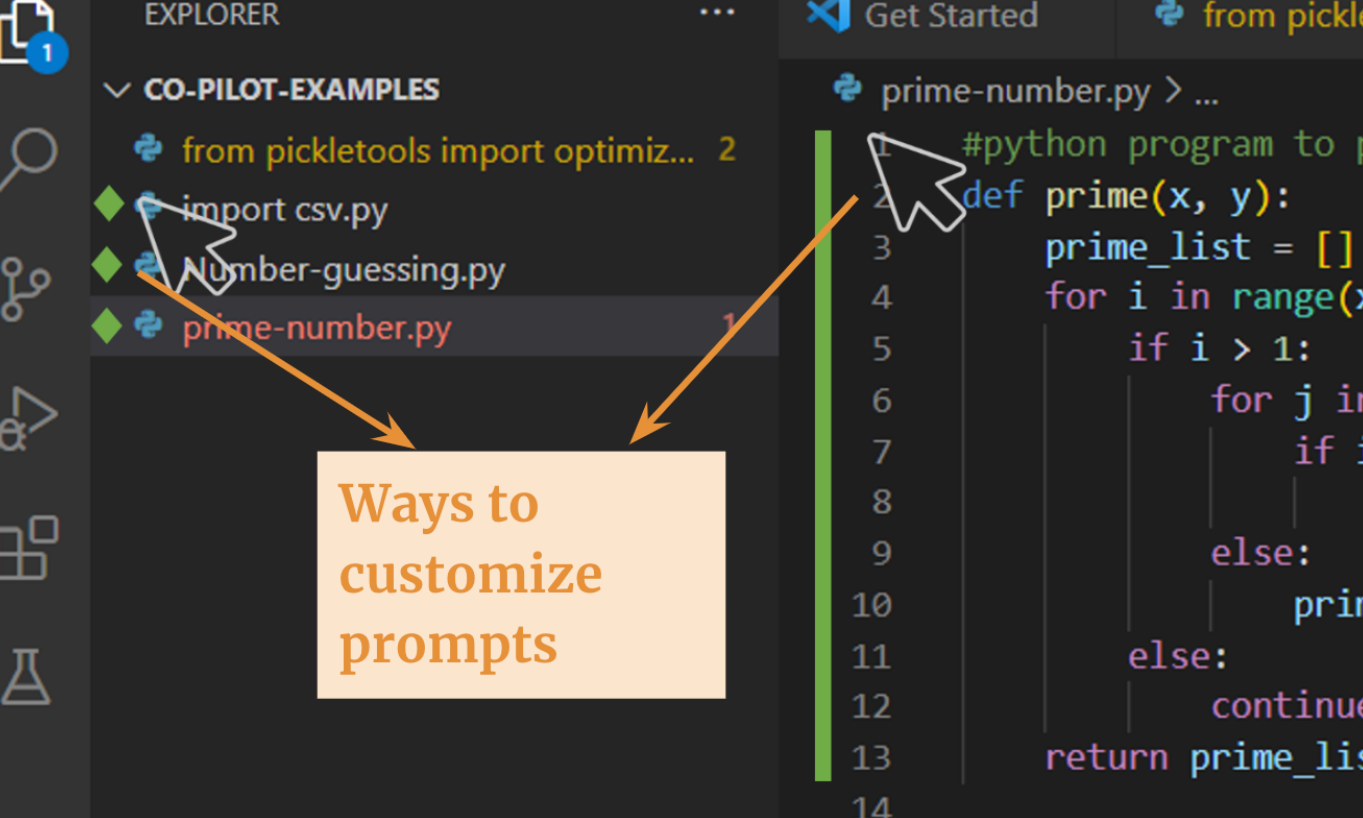}
         \caption{Sliders that enable users to select contexts for code generation}
         \Description{Context slider}
         \label{fig:context}
     \end{subfigure}
     \caption{Two control mechanisms that allow users to communicate intentions to the AI tool. (a) control panel allows users to select system roles at the project initialization; (b) allows users to adapt AI behavior during the programming sessions.  }
     \label{fig:control_overall}
\end{figure}

Existing AI code generation tools require developers to include information in the code they are working on to produce desired AI outputs. However, this interaction paradigm makes it challenging for developers to communicate their intentions and thus evaluate AI tool's benevolence (§~\ref{sec:study1_challenge_prompting}). Therefore, it is crucial to provide developers effective ways to \textbf{convey short-term and long-term goals and preferences}. To help bridge this communication gap, we designed two mechanisms for developers to indicate intention (§~\ref{sec:study1_assess_benevolence}) and preferences for AI's approach (§~\ref{sec:study1_assess_integrity}) when generating suggestions. To complement the existing natural language interface, we designed control mechanisms in graphical interfaces. Specifically, we designed a control panel (Figure \ref{fig:control}) that enables developers to set explicit intentions and define goals for using the AI tool at the project initialization. In the control panel, developers can specify specific benefits they expect to gain from using the AI tool in the programming sessions (e.g., to help them speed up by serving as a prototyping tool or to help them learn as a programming tutor). We chose to use system roles as metaphors since it has been shown to effectively bridge the communication gap between users and large language models~\cite{shenChatGPTWeTrust2023}. Users can also further customize settings by adjusting the configuration on the right side of the control panel. We included options such as suggestion scope, the maximum length of suggestion, the timing of suggestion, and the type of validation (e.g., only suggest solutions that pass security checks). We also designed a context adjustment slider (Figure~\ref{fig:context}) that enables developers to adapt AI behavior further during the programming sessions. Users can drag the control bar next to each file name or the code snippets to manually select the context they would like to include as part of the prompts for code generation.

% P9 wanted to have a ``\textit{sensei}'' version of Copilot, which is more ``\textit{endearing}'' and would ``\textit{invest in you by suggesting what you could learn}'' based on their programming experiences, but find a lack of signals that AI would prioritize their leanings instead of simply outputting suggestions. Similarly, P5 wished to have more guidance and feedback on strategies to harness AI effectively in their workflow. 

\subsection{Study procedure, participants, and data analysis}
We conducted one-to-one 60-minute design probe sessions with 12 developers with diverse programming experience and experience with AI code generation tools from social media and a large technology company. To recruit participants, we emailed 600 randomly selected developers and advertised on social media. We selected participants with various levels of experience with AI tools while ensuring diversity in race, age, and work experience. We stopped recruiting after hearing repeating themes in the interviews. Our final sample includes nine males and three females from different racial groups whose programming experience ranges from 4 to 45 years. All participants in Study 2 have experience with Copilot - 8 use it regularly, 2 recently started using it, and two have used it but are no longer using it.
Detailed profiles of our participants are presented in the Appendix (Table \ref{tab:study2-participant}). To capture a broader range of experiences, we didn't invite participants in Study 1 to participate in Study 2 again. 

The co-design session starts with brief questions on developers' trust attitudes toward AI code generation tools. We then showed the three sets of design concepts to the participants. The visual representations of the design concepts were presented in Microsoft PowerPoint. Each concept is animated to show a sequence of actions to demonstrate the interaction. During the session, we explained each design and asked for participant feedback and reactions, including questions on how they imagined using the proposed features in real life and if and how the features contribute to trust. We also encouraged participants to brainstorm new features. The study procedures received approval from the Institution Review Board. 

Similar to Study 1, all sessions were video and audio recorded and later transcribed. The data analysis followed the procedure of deductive thematic analysis~\cite{braunReflectingReflexiveThematic2019}, following the structure of each design concept. In the analysis, we focused on analyzing ways that interface features are helpful or not helpful for participants to evaluate the trustworthiness of the AI tool, especially the factors identified in Study 1. We also looked for potential risks and places of improvement for each design concept.

\subsection{Study 2 findings}
\subsubsection{Demonstrating AI's practical benefits via usage statistics}
Participants found that the explicit information on Copilot's abilities in both overall and situational usage dashboards was helpful for aligning their expectations with AI's ability. For example, P8 thought that ``\textit{the aggregated measures of how I've used Copilot over time helps me form an image of my relationship with Copilot, which helps me evaluate Copilot's performance and form informed goals.}'' 
% P4 appreciated the ability to see how many suggestions have been made, which helps them access how much Copilot has helped them. 
Several participants (P5, P6, P10) agree that statistics such as suggestion acceptance rate and time saved are useful for demonstrating the AI tool's practical benefit and can help them calibrate their trust in it. For example, P10 suggests: ``\textit{if I can see a quantifiable number of how much Copilot increases my productivity or saved me time, I'm more prone to depend on it more.}'' While we included file-level statistics to help developers calibrate their trust for different situations, P8 wished to see more granular breakdowns of Copilot's performance based on functional concepts so that they can better ``\textit{navigate that space with which topics is Copilot the best at}.''

% Besides benefits, participants also find the statistics to be helpful for learning the limits of Copilot. When P7 saw the statistics of a number of edits they've made to Copilot's suggestions, they are reminded ``\textit{the hassle working with it}'' and that it requires them to ``\textit{learn or to adjust in order to make it work}.''  

At the same time, it can be challenging for users to interpret the statistics shown on the dashboard. For example, P9 worried that: ``\textit{I also need to analyze the correlation and causation, the statistical numbers. I think it's just put into many works to developers}.'' In addition to the numbers, P7 prefers more actionable insights: ``\textit{it's the performance of the Copilot, not my performance, so there's nothing that I can change just based on this...to improve working efficiency with the Copilot.}'' 
Similarly, P2 wished that the dashboard could not only tell users ``\textit{how users used things},'' but also ``\textit{how to use something},'' by including some actionable tips on how to use Copilot in unobtrusive ways. In addition, participants were concerned about potential privacy issues, especially for workplace surveillance when tracking telemetry data. For example, P12 worried that organizations would use the tracked data to evaluate employees.
% : ``\textit{It's scary if this feature can be used by those companies in evaluating how good the programmers are. They will just fire those programmers with bad coding skills}.''

\subsubsection{Offering quality indicators to support evaluation of AI suggestions}
Participants found that the quality indicators at different levels are helpful for them in more efficiently and effectively assessing the quality of code suggestions. For example, P2 thought that the file-level confidence indicates the helpfulness of the AI tool in nuanced and accurate ways: ``\textit{If I know the Copilot is not very familiar with this code, I am not going to have high expectations that the code the Copilot produces will be accurate}.'' 
% \begin{quote}
%     ``\textit{If I know the Copilot is not very familiar with this code, I am not going to have high expectations that the code the Copilot produces will be accurate. If however, Copilot says, yeah, I got this, I know this code, I have a higher confidence in the code that Copilot is producing. Understanding how familiar Copilot is with the code of context helps me understand how useful Copilot suggestions are going to be.}'' (P2)
% \end{quote}
P8 thinks that additional transparency helps them make quick and reliable trust judgments: ``\textit{low familiarity can be a sign of vulnerability for the machine. If I know [the AI tool] is not good at it. I will be more vigilant, careful when I'm writing the code myself or incorporating it... [the transparency signals] help me know how much I should be relying on it}.''
% \begin{quote}
%     ``\textit{When I say low familiarity, it's a sign of vulnerability for the machine. I know this thing is not good at it. I will be more vigilant, careful when I'm writing the code myself or incorporating it... it helps me know how much I should be relying on it.}'' (P8)
% \end{quote}
%
These signals also help prompt subsequent user actions, helping developers integrate AI suggestions into their workflow.
%P1 thought that the function level and token level confidence indicators would allow them to make better judgments by pointing out potential fixes: ``\textit{knowing what's the confidence score, would tell me like pay more attention or not to what is being proposed here. Scrutinize it more or not.}'' Similarly, 
P5 uses the highlights of low-confidence tokens to guide their validation process and ``\textit{target where I'm reviewing the logic and say, yeah, it wasn't super confident about these parts, so I should look more closely at what it did there}.'' 

% Many users reported that explanations can be explicit indicators of whether the particular scenario is a good use case for Copilot and help build approriate expectation to Copilot's capability in different situations over time. For others, explanations shows copilot's vulnerability and help demystifying copilot (e.g., demonstrate that copilot does not hide when it doesn't know). 

%At the same time, the explanations of Copilot confidence can also shape developers' general perception over time. For example, P7 explained that the transparency signals could help them understand potential errors that Copilot can make: ``\textit{seeing the mistakes that it can potentially make helps me realize it's less mysterious because you can see the potential mistake or the principles that you can fall into}''
% \begin{quote}
%     ``\textit{Actually presenting or these confidence scores makes me feel that it is less intelligent because it provides...it was more about seeing the mistakes that it can potentially make that helps me realize it's less mysterious compared to, you can see the potential mistake or the principles that you can fall into}''(P7)
% \end{quote}

% \subsubsection{Guide users actions}
% Explanations can guide users’ actions to better collaborate with copilot (e.g., prioritizing solutions, edit suggestions)

%\paragraph{Potential risks in design}
At the same time, developers indicated that the transparency signals could be hard to interpret without additional contexts. For example, P5 thought that the solution-level confidence indicators were not very helpful because: ``\textit{Even that 20 percent, maybe I have to tweak five lines, it's still a win}.'' P7 also expressed a similar reluctance to fully rely on the numeric metrics: ``\textit{I will pick that a solution even though the confidence score is a little bit lower than the others because it meets my needs better. Human judgment actually knows that that is a better solution}.'' Indeed, many developers also reported challenges in interpreting the context of model confidence numbers---the same numeric score could communicate different information for different developers in a variety of scenarios. Lastly, explicit indications of model confidence also introduced potential bias in users' trust judgments, as users may be ``\textit{more likely to accept without critically thinking about a suggestion}'' or ``\textit{reject a valid solution or a valid suggestion based on low familiarity, even though it's a perfectly valid solution that is ultimately productive}.'' (P6) 

\subsubsection{Communicating developers' intention with control mechanisms}
Participants found the control panel at project initialization and the context adjustment sliders during programming sessions helpful for aligning AI tools with their specific intentions and preferences. The context adjustment sliders offer ``\textit{more tools to guide Copilot to the right answer}'' and allow developers to:
% Instead of me being the student and the model being this imperfect teacher, it's like switching roles. 
``\textit{teach the model what to do for me when I need it}.'' (P1)  The control panel, on the other hand, allows developers to customize how much and what kind of help they get from Copilot at project initialization, which makes the AI more predictable and controllable. For example, P10 once worried that Copilot might introduce unnoticed security bugs, but the option in the control panel for users to customize the type of suggestions could allow them to ``\textit{only get suggestions that have been scanned for any security vulnerabilities}.'' Indeed, control mechanisms allowed users to customize a more reliable and helpful version of Copilot, as P7 described, ``\textit{if the performance is not reliable anymore or if there are suggestions that I don't need, I would turn those off those function just for precision and clarity}.'' Trust was fostered in the process since users felt they had control over what and how the AI will make suggestions: 
% \begin{quote}
``\textit{the ability to set a boundary [for Copilot] and have it respect that boundary is the core of building trust. If it can work in that boundary, then you trust it more, and you can give it more permission.}'' (P4)
% \end{quote}
% like with friendships even. Like, hey, you're great within this boundary, I said, I'll give you more access to this type situation.
 %The control settings allow P7 to narrow down the scope of Copilot to functions that \textit{actually help me saving time}. Similarly, P9 thought that the control settings such as ``only suggesting syntactically correct code'' can help Copilot ``\textit{avoid making silly mistakes}''.

% \subsubsection{By allowing users to have more concrete expectation for copilot}
Interestingly, although we did not explicitly design the control mechanisms to inform users' expectations of AI's abilities, developers thought the control mechanisms allowed them to develop more concrete expectations of what AI can and cannot do. For example, P5 thought that seeing all possibilities to control is almost like interactive documentation for the AI tools' functionalities, showing the full capacity of AI tools. P12 thought that the control panel is especially helpful in project initialization because it allows them to have ``\textit{concrete expectations of what is going to happen},'' such as ``\textit{how many lines of code there will be in suggestions}.'' Others imagined experimenting with functionalities using the controls to understand the strengths and limitations of Copilot in more targeted ways. For example, P7 imagined themselves to ``\textit{turn off everything and see what each function does and see which functions are more helpful},'' which allowed them to have ``\textit{the full scope of what Copilot does}.'' %P9 think that the control settings can help them experiment with Copilot's functionality interactively and on smaller scopes each time, instead of ``\textit{previously on basically the whole scope of code generations}'': ``\textit{Let's say if I only want Copilot to generate variables and some methods like function signatures because I'm really bad at naming things. I give it a try and I found it generated really good names in terms of variable and function signature, then I will stay with them.}'' 

%\paragraph{Potential risks in design}
At the same time, developers expressed concern that too much control could be a burden for users. For example, P5 expressed doubts about the usefulness of the context slider due to its high interaction cost: ``\textit{if I start spending a bunch of time managing what context it has that the utility starts dropping because I'm investing more time than am I getting anything more out of it}.'' The choice of what type of controls to grant users and how to foreshadow their impact on AI behaviors also needs careful consideration. A few developers were confused about some of the current designs, 
%For example, P9 said that: ``\textit{Even though you have something checked or stay or not stay, but you still need to try it out to see if it aligns with your expectations. You don't know what's going to happen here.}'' 
and hoped to see more examples in action and ``\textit{visual cues for what this looks in the editor}'' (P5) or ``\textit{examples like how the code will be different, like turning it on and off}'' (P7), on top of the textual description of the control mechanisms. P10 also thought the presets could be helpful, ``\textit{especially for someone who has no idea about all these customization settings.}

\section{Discussion}
\subsection{Trust in generative AI tools}
Building on prior work that calls for real-world empirical studies of users' trust in AI tools~\cite{kimHumansAIContext2023}, our work contributes a detailed account of users' notions of trust in AI code generation tools based on retrospective interviews with developers who have used such tools in real-life scenarios. Aligning with theories in existing literature~\cite{mayerIntegrativeModelOrganizational1995a, liaoDesigningResponsibleTrust2022}, we observed that developers' trust attitude in AI-powered code generation tool is informed by the tool's perceived practical benefits (\emph{ability}), alignment
with developers' goals (\emph{benevolence}), trustworthy processes (\emph{integrity}) and \emph{situational factors}, such as stakes of the use scenario and the complexity of the programming task.  This echoes prior work indicating that trust is evolving over time~\cite{hollidayUserTrustIntelligent2016}, is situational~\cite{zhangShiftingTrustExamining2022, hoffmanTaxonomyEmergentTrusting2017, jacoviFormalizingTrustArtificial2021a, leeTrustAutomationDesigning2004} and affected by social and organizational contexts~\cite{kimHumansAIContext2023, widderTrustCollaborativeAutomation2021}.

Responding to recent calls to understand how cues in the design of system interface (i.e., trust affordance) communicate the internal trustworthy characteristics of AI to users~\cite{liaoDesigningResponsibleTrust2022}, we observed a lack of trust affordances that can effectively convey the trustworthiness of AI-powered code generation tools. As a result, developers are forced to rely on intuitions accumulated from their limited personal experiences to make trust judgments, which can be inefficient and ineffective and lead to biased trust attitudes. Although our data focused on developers' challenges with AI code generation tools, the challenges of evaluating AI output~\cite{zieglerProductivityAssessmentNeural2022} and conveying goals and intentions to AI using natural language are also observed in other applications of large language models~\cite{guzdialFriendCollaboratorStudent2019, ma2023beyond, zamfirescu2023johnny}. Our work highlights that these challenges not only manifest as usability problems but also affect users' judgment of the trustworthiness of generative AI applications, leading to a potential overreliance on AI or preventing users from taking full advantage of AI. Our work also shows that graphical user interface (GUI) remains crucial in assisting users in establishing  \emph{calibrated} and {warranted} trust in AI, despite recent debates on the possibility of replacing the conventional GUI with the emerging language user interface (e.g.,~\cite{wang2023enabling}).
% voice and other modality, all should consider trust, output itself is not enough, all modality is true. 

% However, we also noticed additional challenges when designing for generative AI. For example, users can easily lose patience once they see errors in both AI-assisted decision-making systems~\cite{dietvorst2015algorithm} and in generative AI systems. The bias can be more damaging since it usually takes time for users to adjust to LLM to get their desired outcome. 

% which situation to consider, give examples, specific situation need to depend 
% need to relieaze that it's sitaution

\subsection{Design for trust affordances in AI code generation tools}
Findings from our design probe study (study 2) additionally shed light on opportunities to support users in building and adjusting their trust in AI tools by augmenting existing interfaces with trust affordances. We outline specific design implications below. While the specific recommendations are derived from the context of AI code generation tools, we believe our advice can also be useful for supporting users in building and calibrating trust with generative AI applications more broadly.

\subsubsection{Encourage structured reflection on AI tool's performance and applicability in specific contexts.}
Developers' trust attitudes are often informed by intuitions accumulated from their personal experiences with AI tools, which can lead to bias and inefficiencies in calibrating their trust in different situations. This suggests a more \textbf{structured approach to align users' expectations by explicitly communicating AI tools' performance and applicability in specific contexts, while also encouraging users' to reflect on the gap between their perception and the tool's actual performance}. In study 2, we evaluated a feedback analytic dashboard that shows personalized statistics of AI tools' performance in different contexts (Figure \ref{fig:Overall usage stats} and~\ref{fig:Situational usage stats}), which proved to be effective in helping developers to form accurate expectations and understand the tool's utility. However, we noticed that simply showing comparisons of statistics might not be enough to prompt users to engage in a reflection, as they can be hard to interpret and require certain data literacy. Therefore, further systems could consider providing more explicit guidance on how users should adjust their trust attitudes or including actionable suggestions on how users can effectively engage with AI tools (e.g., tips on when to use the tool).
% add work about how discussion or reflection could help align expectations

\subsubsection{Support evaluation of AI output using context-aware quality indicators}
Findings from study 1 show that while evaluating AI output forms the basis of trust attitude, developers rely on native methods such as eye-browsing or running the program, which is time-consuming and ineffective, calling for the need to provide \textbf{in-context support for developers to make quick and accurate evaluations of AI output}. In study 2, we explored the potential of three levels of model confidence scores of AI suggestions: token level, solution level, and file level (Figure \ref{fig:local_transparency} and~\ref{fig:file_transparency}). While developers find the confidence indicators useful for evaluating solutions and guiding their actions, there's a clear need to customize these quality indicators to suit diverse preferences and requirements (e.g., explainability or accessibility requirements). One possible design is to allow users to define or adjust metrics based on their specific needs and contexts. The quality indicators could also go beyond explanations of modal mechanisms to include social transparency, such as acceptability of the solution in the community~\cite{chengItWouldWork2023}. Lastly, as previous research on confidence scores has suggested, the design should be wary of users' overreliance~\cite{ agarwalQualityEstimationInterpretability2021}. To mitigate this, it's vital to present quality indicators as part of a broader evaluation framework that includes clear explanations of their meaning and appropriate use. For instance, rather than solely relying on numerical confidence scores, which can be misleading or hard to interpret, AI tools could explain why certain parts of the code were flagged as low confidence to encourage critical reasoning. 

\subsubsection{Afford users to convey short- and long-term goals and preferences}
The various ways that AI tools can be used make it important to help users communicate their intentions clearly. In the design probe study, we demonstrate examples of control options that allow users to customize the timing, characteristics, as well as local context of AI suggestions (Figure \ref{fig:control}). These means of control allow AI tools to better align with users' goals and intentions, communicating the benevolence of the system. This also echoes prior research in the context of AI-powered music generation which indicated that enabling users to steer AI behavior increases trust in AI~\cite{louieNoviceAIMusicCoCreation2020}. However, more controls come with more responsibilities. Designers of generative AI systems need to be cautious about overburdening users with decisions that they are not confident in making or less important to their experience. We suggest that control mechanisms should prioritize places where users have discrepancies or group options and provide users with the option to have simple defaults. We explored persona as a grouping mechanism, which proved helpful. Further systems could also imagine other ways to group them, such as stake of tasks or expertise of users. It's also important to consider how to explain and help users preview the outcome of different control options. Although the users reacted positively to the design probes, they also pointed to the challenges of understanding the control options. Future systems can explore how to introduce the control options more clearly. For example, an interactive onboarding session could potentially address the issue by demonstrating to users the effect of control options in action. Toolsmiths can even consider rolling out at an incremental, progressive clarity on what control means.

\subsection{Limitations and future work}
In this study, we investigated developers' trust in AI-powered code generation tools via qualitative interviews. Future research can build on our qualitative investigation by implementing and evaluating interactive prototypes in controlled experiments to better quantify the effects of interface design on users' trust.
% In addition, our sample was not balanced in gender due to the limitation of sampling via social media and mailing lists.  
% our participants were sampled through social media or email postings in both studies. 

In addition, although we try to reach a diverse population in terms of demographic factors, our sample is still heavily skewed toward male developers, given the general demographics of software engineering. The skewed gender distribution might have affected our findings, given prior research showing that women and minority groups might have different preferences in programming activities (e.g., ~\cite{burnett2010gender}). We call for future work to gain a more in-depth understanding of how female and gender minority developers approach trust in AI tools. 

% Similarly, although we aimed to collect opinions on various AI-code powered code generation tools, our data is heavily biased toward users' experience with Copilot, given its wide adoption among the programmer community. 
Further, our data in Study 1 were collected at a single company. Although we encouraged participants to also discuss their experience outside of the work in Study 1 and intentionally sampled outside of the organization in Study 2, there may be additional needs that we miss because of the specific organizational setting.

Lastly, we collected our interview data between July and August 2022, at a time when AI-powered code generation tools were just starting to emerge. Since then, the landscape of AI-powered code generation tools has been rapidly changing, with several new tools emerging. Existing tools such as GitHub Copilot also introduced updates such as conversation assistants and content exclusion settings. To better contextualize our findings, we provide a description of the features of GitHub Copilot and Tabnine as of July 2020 in the Appendix. Although the core interaction paradigm of AI suggesting code snippets based on code context and natural language prompts remains unchanged, we encourage future research to explore the effect on trust given the fast-growing adoptions in different communities and organizational settings~\cite{liang2024large}.

% \section{Conclusion}
% AI code generation tools introduce new possibilities but also risks in the context of programming. In this paper, we present a two-stage qualitative study to understand developers' trust in AI code generation tools. In stage 1, we interviewed 17 developers to contextualize developers' trust and identify the main challenges with the current system design in helping developers build appropriate trust. In stage 2, we explored three groups of design concepts to address the above challenges and evaluated them with users. Overall, the study demonstrates the importance of designing system features that can help developers build appropriate trust with AI code generation tools. 

% % Through a two-phase qualitative investigation of developers' trust in AI-code generation tools, we surfaced three groups of users' needs in building appropriate trust in AI tools. We also explored how the current system design can support or fail to support users in building appropriate trust. We showed that AI systems could support users in building appropriate trust by communicating to users about AI's abilities and risks, ways to control AI as well as ways to evaluate AI output. We also designed three groups of UX enhancements to existing AI code generation tools and evaluated them with developers. We discussed design guidelines for future designs of AI code generation tools.  

\begin{acks}
We would like to thank the participants for their valuable insights and
anonymous reviewers for their helpful feedback. We would also like to thank members of the Microsoft Research SAINTES team and members of the University of Washington Social Futures Lab for their thoughtful discussion and feedback. 
\end{acks}

\bibliographystyle{ACM-Reference-Format}
\bibliography{reference}

%% If your work has an appendix, this is the place to put it.
\newpage
\appendix

% Please add the following required packages to your document preamble:
% \usepackage{graphicx}
% \usepackage{lscape}
%\begin{landscape}
\section{Participant information}
\begin{table}[ht]
\centering
\caption{The participants of Study 1. The Column \emph{Exp} indicates the years of programming experience. The \emph{Job Title} was self reported by the participants.}
\label{tab:study1-participant}
\resizebox{1\textwidth}{!}{%
\small\begin{tabular}{l|llllllll}
\textbf{ID} & \textbf{Tool (Frequency)}                                                                                                                                                                                                          & \textbf{Gender} & \textbf{Race}             & \textbf{Age} & \textbf{Education}  & \textbf{Job Title} & \textbf{Exp} \\ \hline
P1           &GitHub Copilot (Daily)                             & Male             & White & 25-34       & Bachelor degree     & Researcher                     & 7                           \\ 
P2          & GitHub Copilot (Weekly)                                                                                                             & Male             & Asian                     & 18-24        & Bachelor degree     & Program Maleager                 & 5                           \\
P3          & GitHub Copilot (Monthly)                                                                                                                                                                                                                           & Male             & White                     & 45-54        & Bachelor degree & Software Engineer                          & 25                            \\
P4          & GitHub Copilot (Monthly), Tabnine (Yearly)                                                                                                                                                                                                                   & Male             & White                     & 25-34        & Bachelor degree     & Software Engineer                & 12                            \\
P5          & GitHub Copilot (Daily), Tabnine (Daily)                                                                                                                                                    & Male             & Asian Indian                     & 25-34        & Ongoing Masters degree     & Software engineer      & 9                            \\
P6          & GitHub Copilot (Daily)                                                                                                                                       & Female             & White & 25-34       & Bachelor degree     & Software Engineer                        & 4                           \\
P7          & GitHub Copilot (Monthly)                                                           & Male             & White  & 25-34        & PhD degree     & Software Engineer                        & 22                           \\
P8          & GitHub Copilot (Weekly)                                                                                                                                                                                                           & Male             & Middle Eastern                     & 35-44        & Bachelor degree       & Security Engineer              & 20                           \\
P9          & GitHub Copilot (Weekly)                                                                                                                                                                                                     & Male             & Asian                     & 25-34        & Master degree       & Software Engineer        & 8                           \\
P10         & GitHub Copilot (Daily)                                                                                                                  & Male           & Black or African American                     & 25-34        & High school diploma       & Software Engineer                          & 8                            \\
P11         & GitHub Copilot (Daily)                                                                                              & Male             & White & 25-34        & Bachelor degree     & Software Engineer           & 9                            \\
P12         & GitHub Copilot (Daily)                                                                                                                                                                                                              & Male             & White                     & 35-44        & Master degree       & Software Engineer                       & 21                           \\
P13         & GitHub Copilot (Daily)                                                                                                                                                   & Male             & White                       & 18-24        & Bachelor degree     & Software engineer                & 6                            \\
P14         & GitHub Copilot (Never)                                                        & Female             & Asian                     & 25-34        & Bachelor degree     & Software engineer                       & 8                           \\
P15         & GitHub Copilot (Daily)                                                                                                                     & Male             & Hispanic or Latino                     & 18-24        & High school diploma          & Software engineer intern               & 3                           \\
P16         & GitHub Copilot (Daily)                                                                                                                                                      & Male           & White                     & 18-24        & High school diploma          & Software engineer intern                    & 2                          \\
P17         & GitHub Copilot (Never)                                                                                                                                                             & Male             & Asian                     & 25-34        & Bachelor degree     & Software engineer                      & 5                           \\
\hline
\end{tabular}%
}%end of \resizebox
\end{table}

\begin{table}[ht]
\centering
\caption{The participants of Study 2. The Column \emph{Exp} indicates the years of programming experience.}
\label{tab:study2-participant}
\resizebox{0.8\textwidth}{!}{%
\small
\begin{tabular}{l|llllll}
\textbf{ID} & \textbf{GitHub Copilot (Frequency)}                           & \textbf{Gender} & \textbf{Race}             & \textbf{Age} & \textbf{Education} & \textbf{Exp} \\ \hline
P1       & I use the tool regularly                   & Female             & White                     & 35-44        & PhD degree    & 4                            \\
P2       & I use the tool regularly                   & Male             & White                     & 55-64        & Bachelor degree      & 45                            \\
P3       & I've tried the tool but no longer using it                   & Male             & Asian                    & 35-44        & Bachelor degree    & 5                            \\
P4       & I use the tool regularly & Male           & Black or African American                     & 25-34        & Bachelor degree      & 6                            \\
P5       & I use the tool regularly          & Male             & White                     & 25-34        & Bachelor degree    & 15                            \\
P6       & I recently started using the tool                   & Male             & White                     & 25-34        & Master degree    & 7                            \\
P7       & I recently started using the tool                  & Female             & Asian                     & 25-34        & Master degree    & 6                           \\
P8       & I use the tool regularly          & Female             & Asian                     & 18-24        & Bachelor degree    & 3                            \\
P9       & I use the tool regularly  & Male             & Asian                     & 25-34        & Master degree      & 6                           \\
P10      & I've tried the tool but no longer using it                   & Male             & Middle Eastern & 18-24        & High school diploma    & 6                            \\
P11      & I use the tool regularly          & Male             & Black or African American                     & 35-44        & Master degree    & 11                           \\
P12      & I use the tool regularly           & Male             & Asian                     & 25-34       & Master degree    & 7                           \\
\hline 
\end{tabular}%
}%end of \resizebox
\end{table}
%\end{landscape}

\section{Study material for Study 1}
\subsection{Example interview questions}
The retrospective interviews were semi-structured, so the questions below only represent a general structure of the interviews. In the actual interviews, we followed up with participants whenever they mentioned topics relevant to their understanding of trust and the challenges they have in building appropriate trust. 
\begin{itemize}
    \item Could you tell us a bit more about the kind of programming project or tasks that you work on? What kind of development activities (e.g, front-end) are you typically involved in? 
    \item What experience do you have with AI-powered code generation tools?
    \item How do you trust the AI tool? 
    \item Can you walk me through the significant moments you collected? 
    \begin{itemize}
        \item What was your task?
        \item How did you interact with the tool? [Feel free to share screen]
        \item How do these interactions affect your trust in the tool? Why?
    \end{itemize}
    \item Now think about your general experience interacting with the tool. How would you define trust?   
    \item Where do you think the trust come from? 
    \item What tasks do you trust/distrust the tool to do? Why? 
    \item Were there moments where you trusted the AI tool but later realized that you shouldn’t? 
    \item Were there moments where you didn’t trust the AI tool but later realized that you should? 
    \item How has your perception of trust in the tool changed over time? 
    \item How would you want to improve the design of  AI-powered code generation tool so that you can trust it more appropriately? 
\end{itemize}

\subsection{Message sent to participants to collect significant moments}
\begin{quote}
    \textit{Hi [Participant Name], Thank you for signing up for the experience in AI-powered code generation tools research study. We would like to invite you for an interview to learn more about your experience. To prepare for the interview, we would like to invite you to collect significant moments in your experience using AI-powered code generation tools (e.g., Copilot) in the next few days. Our goal is for you to collect these significant moments, so that you can reflect on your experience more concretely in the interview. } 
    \textit{Specifically, please aim to share 1 to 3 significant moments each day. Some examples of significant moments are when you are appreciative of, frustrated by, or hesitant/uncertain to use the AI-powered code generation tool (e.g., copilot). For each time you share, you can use one or two sentences to describe the instance, take a screenshot or share a snippet of code. You can share these in our chat directly. We will also send you a quick reminder message everyday morning during the week. In the case that you do not use AI-powered code generation tools during the day, it would also be helpful to share a quick update in the chat (e.g., did not use AI tools today). We will schedule an interview with you after you successfully complete the preparation phase (collect several significant moments).  } 
\end{quote}
\section{Study material for Study 2}
\subsection{Example design probe questions}
We begin the interview by briefing participants that:
\begin{itemize}
    \item We are evaluating the prototype, not you, so feel free to comment on anything.
    \item Do not worry about the technical implementation of the designs. The purpose of the session is to get feedback on the concept of designs, instead of the feasibility of the designs. 
    \item Do not worry about usability (e.g., layout, color, style) of the design
    \item Feel free to think aloud as you look at the design prototypes
    \item The code snippets are only placeholders. Try to imagine how you will use the design in your daily workflow.
\end{itemize}
Next, we ask the following questions to understand participants' understanding of trust.
\begin{itemize}
    \item What experience do you have with AI-powered code generation tools, such as copilot? 
    \item How do you trust the AI tool? 
    \item How do you define trust? 
    \item Are there challenges in knowing what to expect from the AI tool? 
    \item Are there challenges in integrating the AI tool into your workflow? 
    \item How would you want to design the interaction with the AI tool differently so that you can better judge when to trust the AI tool or not?
\end{itemize}    
Present and give brief explanations of the mockups to participants one by one. For each mockup, ask the following questions: 
\begin{itemize}
    \item What do you think of this design? 
    \item How might you use this feature in your daily coding task? 
    \item Thinking about your overall experience interacting with the tool, to what extent do you think it will help you better judge when to trust copilot or not?  
    \item Which of all the mockups is the most helpful in helping you judge when to trust copilot or not? 
    \item What other features do you like to add to this prototype?  
    \item What other features do you like to remove or change to this prototype?  
    
\end{itemize}

\section{Code book for Study 1}
Table \ref{tab:codebook-study1} shows the codebook for the inductive thematic analysis in Study 1. 
% Please add the following required packages to your document preamble:
% \usepackage{graphicx}
% \usepackage{lscape}

% Please add the following required packages to your document preamble:
% \usepackage{graphicx}
% \usepackage{lscape}
% \begin{landscape}
\begin{table}[h]
\centering
\resizebox{\textwidth}{!}{%
\begin{tabular}{|l|l|}
\hline
Category & Code                              \\ \hline
How developers use AI tools              & understand AI tool's utility over time                           \\ \hline
                  & assign different roles of AI tools          \\ \hline
                  & expect different scope of suggestions       \\ \hline
                  & validate before accepting suggestions       \\ \hline
                  & willing to accept without validation        \\ \hline
Factors affecting trust in AI tools      & general trust perceptions                                        \\ \hline
                  & compare AI tools to human                   \\ \hline
                  & effect on productivity                      \\ \hline
                  & stability of performance                    \\ \hline
                  & (mis)aligned expectation on AI tools        \\ \hline
                  & ability to convey intention                 \\ \hline
                  & reliability of suggestions                  \\ \hline
                  & concerns around privacy and security        \\ \hline
                  & transparency of model mechanism             \\ \hline
                  & trust varied by complexity of task    \\ \hline
                                         & trust varied by the granularity of expected suggestion         \\ \hline
                  & trust varied by programming language  \\ \hline
                  & trust varied by stake of task         \\ \hline
                  & trust varied by individual factors          \\ \hline
Evaluating specific AI suggestions       & local judgement differ from global trust perception             \\ \hline
                  & global trust affects local judgement        \\ \hline
                                         & knowing the exact context help evaluate AI suggestions           \\ \hline
                  & explanation help evaluate AI suggestions    \\ \hline
Challenges in building trust in AI tools & lack of support in onboarding experience                         \\ \hline
                  & trust perception shift over time            \\ \hline
                  & initial expectation affects trust building             \\ \hline
                  & build trust via intentional experimentation \\ \hline
                  & prior knowledge shapes trust perception      \\ \hline
                  & success and failure cases shape trust       \\ \hline
                                         & trial and error to build trust in AI \\ \hline
                  & want to understand the limits of AI tools   \\ \hline
                                         & evaluate suggestion based on external references                 \\ \hline
                  & fixing AI's error affects trust             \\ \hline
                  & challenges in validation                    \\ \hline
                  & learning how to control AI tools            \\ \hline
                  & assign too much responsibility on AI tools  \\ \hline
                  & integrate AI tools in workflow              \\ \hline
                                         & expect the AI tools' performance to grow over time               \\ \hline
                  & develop folk theory of how AI works         \\ \hline
\end{tabular}%
}
\caption{Codebook for inductive thematic analysis in Study 1}
\label{tab:codebook-study1}
\end{table}
% \end{landscape}
\section{Code book for Study 2}
Table \ref{tab:codebook-study2} shows the codebook for the thematic analysis in Study 2. 
% Please add the following required packages to your document preamble:
% \usepackage{graphicx}
% \usepackage[normalem]{ulem}
% \useunder{\uline}{\ul}{}
% \usepackage{lscape}
% \begin{landscape}
\begin{table}[h]
\centering
\resizebox{\textwidth}{!}{%
\begin{tabular}{|l|l|}
\hline
Design concept             & Code                                                        \\ \hline
Usage statistics dashboard & demonstrate values of AI tools                              \\ \hline
                           & support exploration of AI capabilities                      \\ \hline
                           & help understand the limitation of AI tools                  \\ \hline
                           & privacy concern around behavior analytics                   \\ \hline
                           & high effort to interpret the stats                          \\ \hline
Quality indicators         & help guide decisions on whether to accept suggestion or not \\ \hline
                           & help show vulnerability and demystifying AI tools           \\ \hline
                           & helpful signals to make trust judgement                     \\ \hline
                           & difficult to interpret numbers                              \\ \hline
                           & potential to introduce bias                                 \\ \hline
Control mechanisms         & help set boundaries and align intentions                    \\ \hline
                           & help build expectations on AI tools                         \\ \hline
                           & settings are hard to understand                             \\ \hline
                           & additional effort of using AI tools                         \\ \hline
\end{tabular}%
}
\caption{Codebook for thematic analysis in Study 2}
\label{tab:codebook-study2}
\end{table}
% \end{landscape}
\section{Features of Github Copilot and Tabnine as of July 2022}
\begin{figure*}[h]
     \centering
         \includegraphics[width=\textwidth]{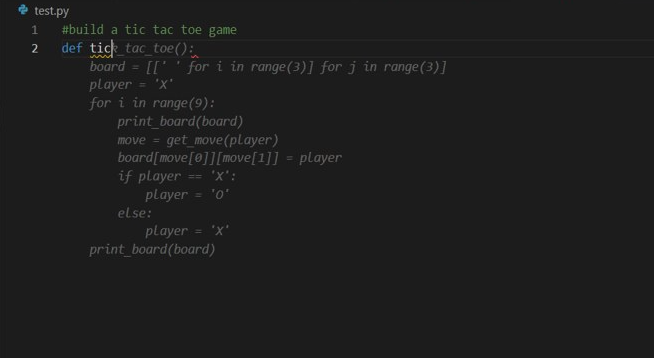}
         \caption{GitHub Copilot interface, as of July 2022}
         \Description{GitHub Copilot interface, as of July 2022, showing a screenshot of an IDE with inline suggestion from Copilot}
         \label{fig:copilot2022}
\end{figure*}

As of July 2022, GitHub Copilot was an AI-powered code generation tool that is integrated into code editors as shown in Figure~\ref{fig:copilot2022}. Based on the official website image of July 2022~\footnote{\url{https://web.archive.org/web/20220701014741/https://github.com/features/copilot} --Retrieval date: 05/02/2024}, GitHub Copilot ``\textit{uses the OpenAI Codex to suggest code and entire functions in real-time, right from your editor.}'' It can generate whole lines or blocks of code based on the comments and preceding code snippets. Copilot also supports multiple programming languages and frameworks, including Python, JavaScript, etc. However, users cannot chat with the tool. Moreover, GitHub Copilot Chat, which allows users to interact with GitHub Copilot to ask and receive answers to coding-related questions, was not available. Features allowing users to select a snippet of code and ask natural language questions~\footnote{\url{https://code.visualstudio.com/docs/copilot/overview} --Retrieval date: 05/02/2024} also became available after our study. Similarly, Tabnine also only supported code completion within editors in July 2022~\footnote{\url{https://web.archive.org/web/20220705023816/https://www.tabnine.com/} --Retrieval date: 05/02/2024} and only supported a chat interface after our study.

\section{Design concepts shown in the study}~\label{fig:design_concepts_appendix}
\begin{figure*}[t]
     \centering
     \begin{subfigure}[b]{\textwidth}
         \centering
         \includegraphics[width=\textwidth]{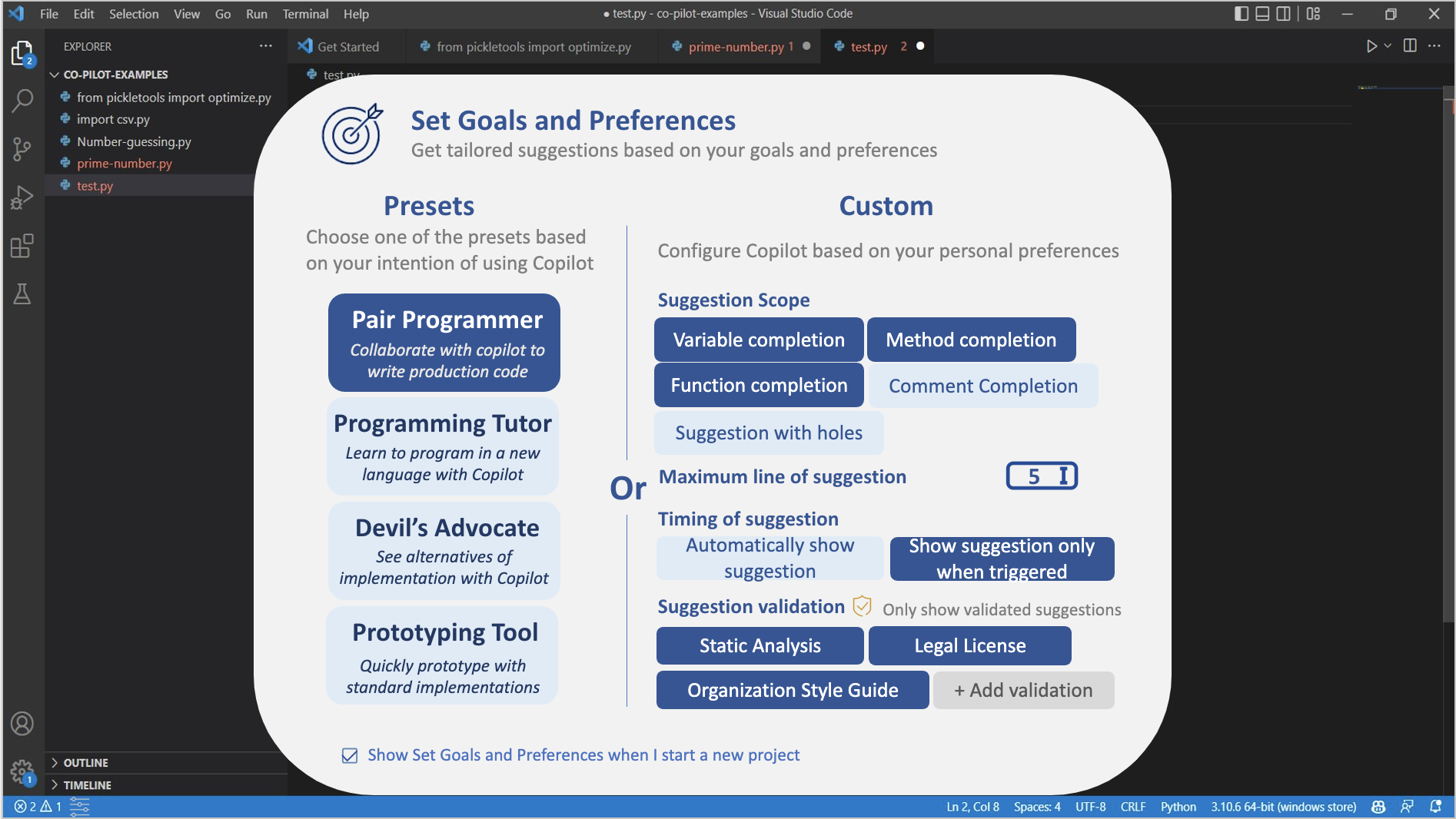}
         \caption{Control panel}
         \Description{ }
     \end{subfigure}
    \newline
     \begin{subfigure}[b]{\textwidth}
         \centering
         \includegraphics[width=\textwidth]{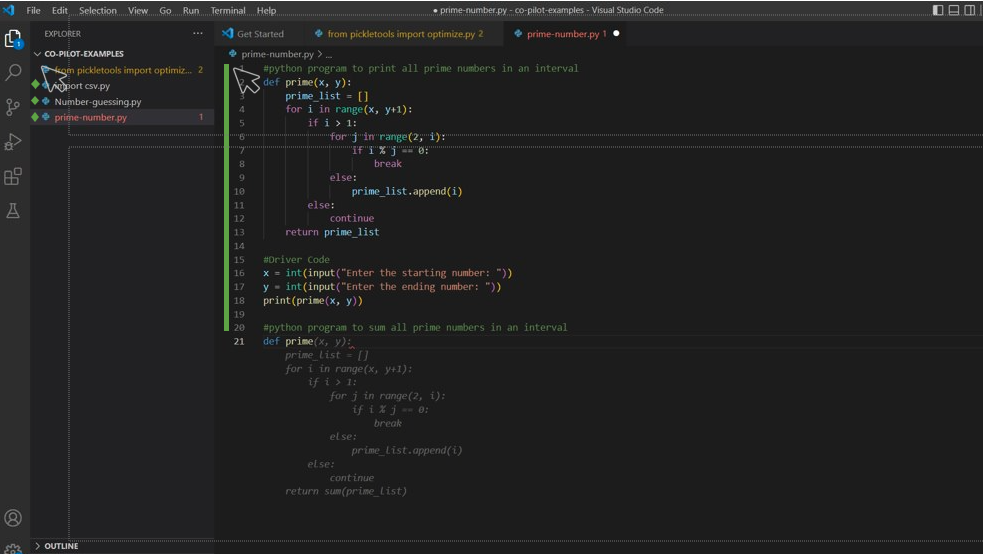}
         \caption{Context slider}
         \Description{Context slider}
     \end{subfigure}
     \caption{Group 1: Control mechanisms}
     \label{fig:control-appendix}
\end{figure*}

\begin{figure*}[t]

 \begin{subfigure}[b]{\textwidth}
         \centering
         \includegraphics[width=\textwidth]{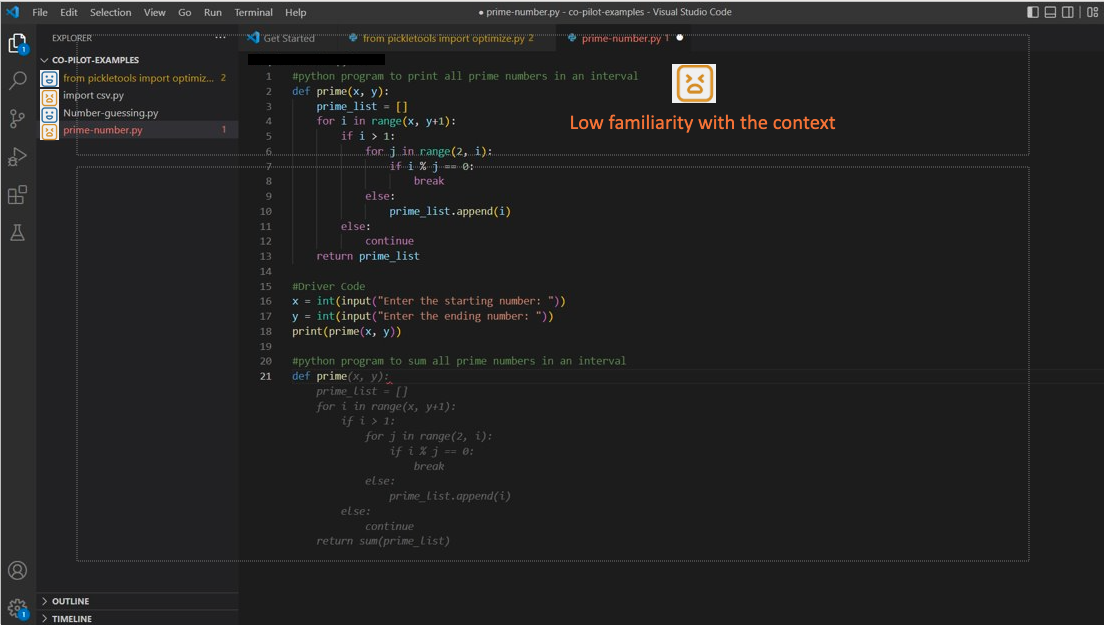}
         \caption{File-level familiarity explanation}
         \Description{File-level familiarity explanation}
         \label{fig:file_transparency_appendix}
\end{subfigure}

 \begin{subfigure}[b]{\textwidth}
    
         \includegraphics[width=\textwidth]{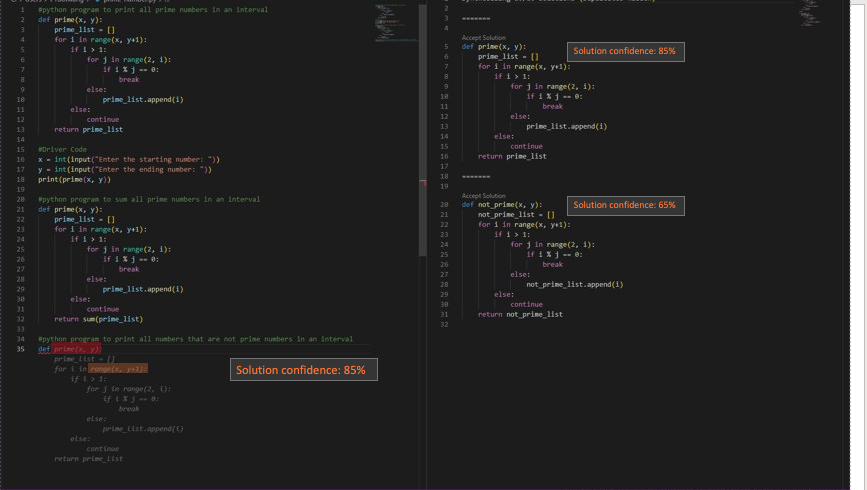}
         \caption{Solution-level and token-level confidence explanations}
         \Description{Solution-level and token-level confidence explanations}
         \label{fig:local_transparency_appendix}
\end{subfigure}
\caption{Group 2: Quality indicators of AI suggestions}
\label{fig:transparency-appendix}
\end{figure*}

\begin{figure*}[t]
    
         \includegraphics[width=\textwidth]{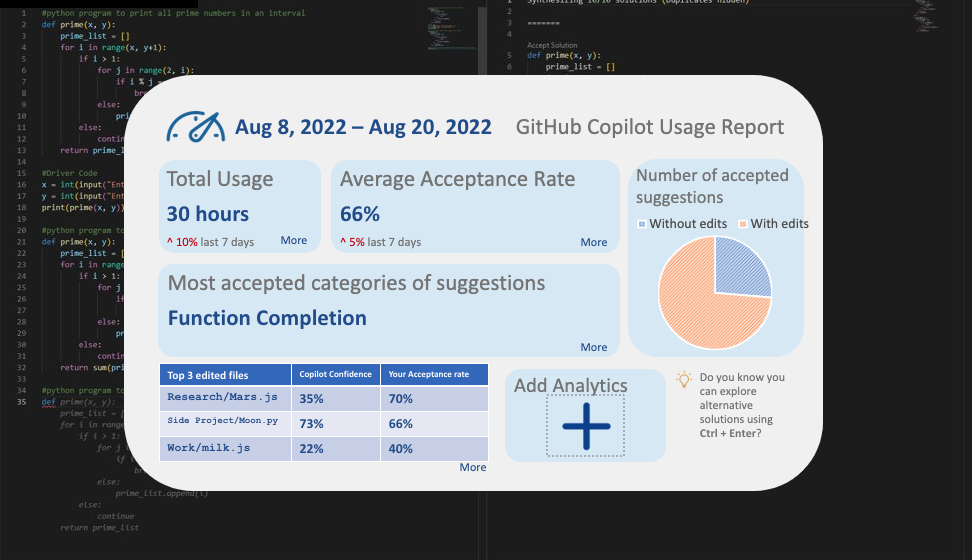}
         \caption{Group 3: Usage statistics dashboard}
         \Description{}
         \label{fig:analytics}
     \label{fig:design_repo}
\end{figure*}

In Figure~\ref{fig:control-appendix},~\ref{fig:transparency-appendix},~\ref{fig:design_repo}, we show the design prototypes that we showed to the study participants.

\end{document}